\documentclass[aps,preprint]{revtex4}
\usepackage{amsfonts}
\usepackage{amsmath}
\usepackage{amssymb}
\usepackage{mathrsfs}
\usepackage{subfigure}
\usepackage{graphicx}
\usepackage{epstopdf}
\usepackage{float}
\usepackage{color}
\usepackage{bm}
\usepackage{ulem}
\usepackage{xcolor}
\usepackage{appendix}
\usepackage{booktabs}
\usepackage{hyperref}
\usepackage{array}

\begin{document}
\title{ Lyapunov Exponents and Phase Transitions of Born-Infeld AdS Black Holes}
\preprint{CTP-SCU/2023005}
\author{Shaojie Yang$^{a}$}
\email{yangshaojie@stu.scu.edu.cn}
\author{Jun Tao$^{a}$}
\email{taojun@scu.edu.cn}
\author{Benrong Mu$^{b}$}
\email{benrongmu@cdutcm.edu.cn}
\author{Aoyun He$^{a}$}
\email{heaoyun@stu.scu.edu.cn}
\affiliation{$^{a}$Center for Theoretical Physics, College of Physics, Sichuan University, Chengdu, 610065, China}
\affiliation{$^{b}$Physics Teaching and Research section, College of Medical Technology, Chengdu University of Traditional Chinese Medicine, Chengdu 611137, China}
\begin{abstract}
In this paper, we characterize the phase transitons of Born-Infeld AdS black holes in terms of Lyapunov exponents. We calculate the Lyapunov exponents for both null and timelike geodesics. It is found that black hole phase transitions can be described by multiple-valued Lyapunov exponents. And its phase diagram can be characterized by Lyapunov exponents and Hawking temperature. Besides, the change of Lyapunov exponents can be considered as order parameter, and exists a critical exponent $1/2$ near critical point. 
\end{abstract}
\maketitle
\section{Introduction}
The study of black hole thermodynamics has led to significant advances in our understanding of fundamental physics, including the interplay between gravity and quantum mechanics. Hawking's area theorem proved that when two black holes merge together, the surface area cannot be smaller than the sum of the surface areas of two black holes \cite{Hawking:1971tu}. Inspired by this property, Bekenstein established the concept of black hole entropy corresponding to its area and put forward the viewpoint of black hole thermodynamics \cite{Bekenstein:1973ur}. Moreover, according to the theory of Hawking radiation \cite{Hawking:1974rv,Davies:1974th,Hawking:1975vcx}, black hole surface gravity can be related to its temperature. Afterwards, Bardeen established the four laws of black hole thermodynamics analogous to the laws of thermodynamics, further maturing this theory \cite{Bardeen:1973gs}. 

Black hole thermodynamics with negative cosmological constant gave rise to a set of new insight to investigate the nature of black holes. One of the important findings from this research is that a Hawking-Page phase transition exists between pure anti-de Sitter (AdS) space and black holes for asymptotically AdS space \cite{Hawking:1982dh}. Subsequent research on the Reissner-Nordström-AdS black hole has revealed that its phase transition behavior bears a formal resemblance to the Van der Waals gas. And a first-order phase transition exists in the canonical ensemble, where the small and large black hole can correspond to the liquid and gaseous phases, respectively \cite{Chamblin:1999tk,Chamblin:1999hg}. For this type of black holes, free energy and heat capacity was used to probe the phase transitions and this topic has been studied in various papers \cite{He:2010zb,Guo:2021enm,He:2016fiz,Yao:2020ftk,Huang:2021iyf,Bai:2022hti,Wang:2019urm,Li:2019dai,Wang:2019kxp,Wang:2018xdz}.

Chaos theory deals with intrinsic models and physical laws of dynamical systems that are disordered and highly sensitive to initial conditions. Black holes are considered to be nonlinear and non-integrable systems, and there have been numerous studies conducted on their chaotic behavior. \cite{Bombelli:1991eg,Suzuki:1996gm,Hartl:2002ig,Lu:2018mpr}. The tachyonic mode of perturbation in RN AdS black holes has been studied \cite{Gubser:2000ec}. The chaotic features of multi-black hole spacetimes were presented on a holographic screen \cite{Hanan:2006uf}. As a useful parameter in study of chaotic systems, Lyapunov exponents $\lambda$ could be used to distinguish different types of orbits for both continuous and discontinuous systems. When $\lambda=0$, the system is in certain stable mode and when $\lambda<0$ the orbit exhibits asymptotic stability which grows better as the absolute value of $\lambda$ increases. For $\lambda>0$, the system is chaotic and unstable, which means that any nearby orbits will diverge eventually via a subtle perturbation. The Lyapunov exponents can be applied to probe the motion of particles for Schwarzschild-Melvin spacetime \cite{Wang:2016wcj} and for accelerating and rotating black holes \cite{Chen:2016tmr}. It has been shown that for spherically symmetric spacetimes, QNMs can be explained as particles trapped in unstable circular null geodesics and slowly leaking out, which can be given by the Lyapunov exponent \cite{Cardoso:2008bp}. Interestingly, black hole QNMs were found to have intense change near the phase transition \cite{Liu:2014gvf,Momennia:2018hsm}. Thus, characteristic phenomenon of QNMs near phase transition points could be reflected by Lyapunov exponents, which reveals that black hole phase transition can be related to the divergence and convergence rate of the particle orbits around the black hole equatorial plane.

By building upon the aforementioned connections between quasinormal modes (QNMs) and black hole phase transitions, it is possible to establish a direct relation between Lyapunov exponents and such transitions \cite{Guo:2022kio}. In this work, the occurrence of a phase transition can be identified by a discontinuous jump in the value of the Lyapunov exponent $\lambda$. Correspond to the phase transition point, the critical exponent of $\lambda$ has been calculated as $1/2$ which is just the same  as the circular orbit radius. And Lyapunov exponents can be used to probe black hole phase transitions both in stable and unstable orbits. This work has opened a new window for the extended black hole spacetimes or the phase space beyond RN AdS black holes.  

Non-linear electrodynamics (NLED) is a generalization of Maxwell theory in high field and micro length scale \cite{Bi:2020vcg}. Coupled with gravity, NLED has significant effects on the black hole in dynamics \cite{Gibbons:1985ac,Bartnik:1988am,Bizon:1990sr,Lavrelashvili:1992ia}. It is proved that in nonlinear electrodynamics photons move along null geodesics in the effective geometry instead of actual spacetime. This means that the motion of light can be viewed as an electromagnetic wave traveling through a classical dispersive media in a nontrivial vacuum \cite{Dittrich:1998fy}. The medium modifies the equations of motion described in nonlinear dynamics, such as effective potential and Lyapunov exponents. Born-Infeld(BI) electrodynamics, as one of the most popular NLED, modifies the traditional electrodynamic laws to obtain the finite value of point charge self-energy by adding another BI parameter \cite{Born:1934gh}. The BI AdS black hole solution in the presence of a cosmological constant was studied in \cite{Dey:2004yt,Cai:2004eh}. The thermodynamic behaviors and phase transitions of BI black holes in different fields have also been investigated \cite{Gunasekaran:2012dq,Bai:2022vmx,He:2022opa,Wen:2022hkv,Tao:2017fsy}. 

In this paper we explore the relationship between Lyapunov exponents and phase structures of BI AdS black holes for the geodesics of both massless and massive particles. The rest of this paper is organized as follows. Thermodynamical quantities of Born-Infeld AdS black holes are briefly reviewed in Sec. II. We discuss the BI AdS black hole phase stuctures and S-L phase transition in Sec. III. Then we study the stability of geodesics and the expressions of Lyapunov exponents in Sec. IV. We show the relationship between Lyapunov exponents and black hole phase transitions in Sec. V. Finally we give a conclusion of the whole paper in Sec. VI.

\section{Thermodynamics Of BI AdS Black Holes}

The Einstein-Hilbert-Born-Infeld action \cite{Born:1934gh} for D-dimensional($D\geqslant 4$)is
\begin{gather}
\label{eq:action}
S=\frac{1}{16\pi}\int{d^Dx\sqrt{-g}(R-2\Lambda+\mathcal L(F))},
\end{gather}
where the cosmological constant $\Lambda$ can be expressed by AdS space radius $l$ as
\begin{gather}
\label{eq:cc}
\Lambda=-\frac{(D-1)(D-2)}{2l^2}.
\end{gather}
The nonlinear electromagnetic term $\mathcal L(F)$ is given by
\begin{gather}
\label{eq:non elec}
\mathcal L(F)=4\beta^2(1-\sqrt{1+\frac{F^{\mu\nu}F_{\mu\nu}}{2\beta^2}}),
\end{gather}
here $\beta$ is the Born-Infeld parameter, which leads to an upper limit of the field strength \cite{Cai:2004eh}. When $\beta\rightarrow\infty$, the action reduces to the Maxwell term
\begin{gather}
\label{eq:maxwell}
\mathcal L(F)=-F^{\mu\nu}F_{\mu\nu}+\mathcal O(\frac{1}{\beta}).
\end{gather}
The metric of a 4-dimensional Born-Infeld AdS black holes is given by \cite{Bi:2020vcg,Wang:2019cax} 
\begin{gather}
\label{eq:metric}
ds^2=-f(r)dt^2+\frac{dr^2}{f(r)}+r^2d\Omega_2^2,
\end{gather}
with
\begin{gather}
\begin{split}
\label{eq:fr}
f(r)&=1-\frac{2M}{r}+\frac{r^2}{l^2}+\frac{2\beta^2r^2}{3}(1-\sqrt{1-z})
+\frac{4q^2}{3r^2}\,{_2F_1}[\frac{1}{4},\frac{1}{2},\frac{5}{4},z],
\end{split}
\end{gather}
where $_2F_1(a,b,c,z)$ is the hypergeometric function, and $z$ satisfies
\begin{gather}
\label{eq:z}
z=-\frac{q^2}{\beta^2r^4_+}.
\end{gather}
The parameters $M$ and $q$ determine the mass and the charge of black holes,
The radius of event horizen $r_+$ is the solution of $f(r_+)=0$ , then the mass $M$ yields
\begin{gather}
\label{eq:MT}
M=r_+\left\{\frac{1}{2}+\frac{r_+^2}{2l^2}+\frac{\beta ^2 r_+^2}{3}(1-\sqrt{1-z})+\frac{2q^2}{3r_+}\,_2F_1[\frac{1}{4},\frac{1}{2},\frac{5}{4},z]\right\},
\end{gather}
and Hawking temperature $T$ are expressed as
\begin{gather}
\label{eq:T}
T=\frac{1}{4\pi}[\frac{1}{r_+}+\frac{3 r_+}{l^2}+2\beta ^2 r_+(1-\sqrt{1-z})].
\end{gather}
The first law of black hole thermodynamics presents the anaolog of thermodynamic expression as \cite{Bekenstein:1973ur} 
\begin{gather}
\label{eq:law}
dM=TdS+\Phi dq,
\end{gather}
where
\begin{gather}
\label{eq:S}
S=\pi r^2_+,
\end{gather}
is the black hole entropy corresponding to its horizon area, and
\begin{gather}
\label{eq:phi}
\Phi=\frac{q}{r_+} {_2F_1[\frac{1}{4},\frac{1}{2},\frac{5}{4},z]},
\end{gather}
is the electrostatic potential conjugated to charge $q$.
By computing the Euclidean action in the semiclassical approximation, the free energy of the black hole satisfies $F=M-TS$, yielding
\begin{gather}
\begin{split}
\label{eq:F}
F=\frac{r_+}{4}-\frac{\beta ^2 r_+^3}{6}-\frac{r_+^3}{4l^2}+\frac{1}{6} \beta ^2 r_+^3 \sqrt{1-z}+\frac{2 q^2 \, _2F_1\left(\frac{1}{4},\frac{1}{2},\frac{5}{4},z\right)}{3 r_+}.
\end{split}
\end{gather}

By dimensional analysis, the powers of $l$ can be used as the physical quantities scale \cite{Guo:2022kio} as
\begin{align}
\tilde{q}=q/l, \quad\tilde{r}_+=r_+/l, \quad\tilde{T}=Tl, \quad\tilde{F}=F/l, \quad{M}=M/l, \quad\tilde{r}=r/l,  
\end{align}
here all tildes are dimensionless quantities. For the convenience of writing, we will directly represent these quantities without displaying the tilde.

\section{Phase transitions of BI AdS Black Holes}
In this section, we investigate the phase transitions of BI AdS black holes. The characteristics of black hole solutions can be captured by critical points in the $T-r_+$ configuration space where $q$ reaches $q_c$. 
The critical points for a given $\beta$ are determined by
\begin{gather}
\label{eq:critical}
\frac{\partial T}{\partial r_+}=0,\quad\frac{\partial^2 T}{\partial r_+^2}=0.
\end{gather}

\begin{table}[t]
\centering
\setlength{\tabcolsep}{5mm}{
\begin{tabular}[b]{c|cccccc}
\toprule[2pt]
$\beta$ & $q_c(s)$ & $q_c(l)$ & $r_{+c}(s)$ & $r_{+c}(l)$ & $T_c(s)$ & $T_c(l)$ \\
\midrule[1pt]
2 & 0.18099 & 0.18293 & 0.26348 & 0.34994 & 0.25708 & 0.25566 \\
5 & 0.08301 & 0.16867 & 0.08438 & 0.40318 & 0.51594 & 0.25944 \\
10 & 0.04216 & 0.16716 & 0.04116 & 0.40704 & 0.65578 & 0.25979 \\
20 & 0.02116 & 0.16679 & 0.02045 & 0.40795 & 0.73756 & 0.25987 \\
50 & 0.00847 & 0.16669 & 0.00817 & 0.40820 & 4.12272 & 0.25990 \\
$\infty$ & $\sim$ &0.16667 & $\sim$ & 0.40825 & $\sim$ & 0.25990 \\
\bottomrule[2pt]
\end{tabular}}
\caption{\label{table1}
$q_c$ and $r_{+c}$, $T_c$ of critical points of the BI AdS black hole with different $\beta$. There is no physical solution (the solution is unreal) of critical point when $\beta<\beta_0=1.903$. When $\beta\rightarrow\infty$ the results reduce to RN AdS black holes.}
\end{table}

Then we can get temperature $T_c$, black hole charge $q_c$ and event horizon radius $r_{+c}$ at the critical point. It is difficult to obtain analytical solutions of critical point for BI AdS black holes, thus we can numerically gather the part solutions of equations in Table. \ref{table1}. We can see there exist two critical points \cite{Gunasekaran:2012dq}, one is at the larger $q_c$ and can be noted as $l$ by $r_{+c}(l)$, $q_c(l)$, $T_c(l)$, while the smaller one noted as $s$. When $q>q_c(l)$ or $q<q_c(s)$, the small and large black hole phases can be continuously converted into another one and no phase transition exists \cite{Chamblin:1999tk}. The phase transition may appear when $q_c(s)<q<q_c(l)$. An important behaviour in Table. \ref{table1} is that as $\beta$ increases, $q_c(s)\rightarrow0$ and $r_{+c}(s)\rightarrow0$. As $\beta\to\infty$, the BI AdS black hole reduces to RN AdS type, and there is only one critical point as $r_{+c}=1/\sqrt{6}$, $q_c=1/6$ and $T_c=1/\pi \sqrt{2/3}$ \cite{Guo:2022kio}. Another interesting point is that as $\beta$ decreases, the gap between $q_c(s)$ and $q_c(l)$ becomes narrow. When $\beta=\beta_0$, these two critical points eventually merge together and the phase transition disappears. Based on the phase transition law, we can calculate $\beta_0$ by requiring 
$\partial^3 T/\partial r_+^3=0$. The result is $r_{+0}=0.312$, $q_0=0.186$, $\beta_0=1.903$. Thus when $\beta<\beta_0=1.903$ the critical point of the black hole vanishes and no phase transition exists.

\begin{figure}[t]
        \centering
        \subfigure[]{\includegraphics[height=5cm]{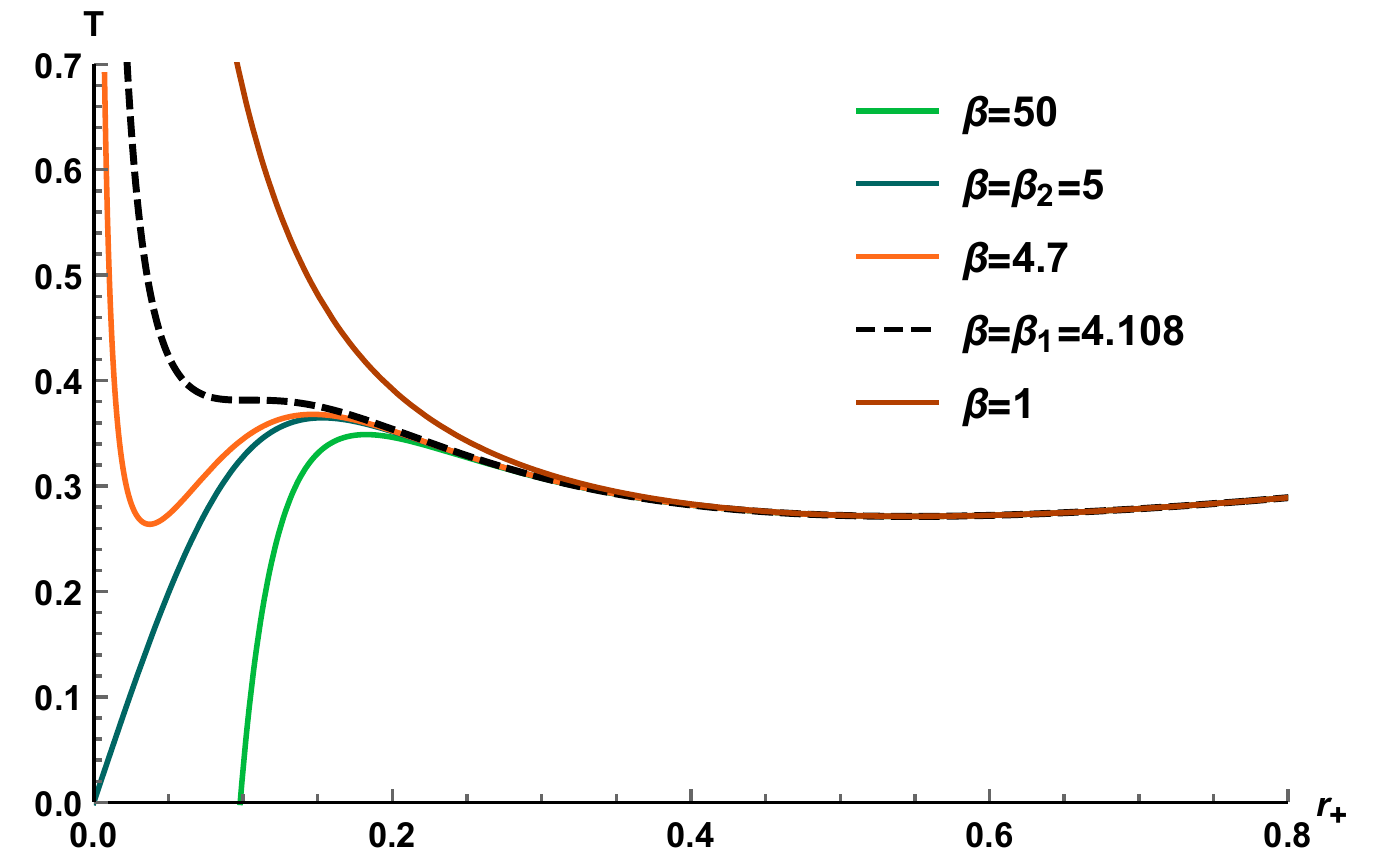}
        \label{fig:fig1}}
        \caption{\label{fig:Fig.1}The curves of Hawking temperature $T(r_+)$ and event horizen $r_+$ with $\beta=50$ (a), $\beta=\beta_2=5$ (b), $\beta=4.7$ (c), $\beta=\beta_1=4.108$ (d) and $\beta=1$ (e) respectively. Curves (a) and (b) have two extreme points, curve (c) has three extreme points. The curve (d) is the situation that number of extreme points change from three to one, indicating the beginning of phase structures nonexistence. For $\beta<\beta_1$ black holes have no phase structure such as  $\beta=1$ (e). In particular, for (b), the curve passes through (0,0) point.}
\end{figure}

For simplicity, we can set $q=0.1$ to show how the phase structures of BI AdS black hole varies with $\beta$ and we depict $T(r_+)$ curve with various $\beta$ in FIG.\ref{fig:Fig.1}.
According to the value of $\beta$, it can be divided into two branches when studying the phase structures of the BI AdS black hole, i.e. Maxwell branch (green part) and BI branch (orange part). Note that when $\beta=\beta_2=(2q)^{-1}$, the Hawking temperature satisfies $\lim\limits_{r_+\to 0}T(r_+)=0$, which can be defined to be the threshold condition of these two branches. For black holes with charge $q=0.1$, we can calculate the $\beta_2=5$. For $\beta>\beta_2$, the phase structures of the BI AdS black holes are qualitatively similar to the RN AdS black holes which have three phases, i.e. small BH, intermediate BH and large BH. And that is why we call that as the Maxwell branch. When $\beta=\beta_1$, the Hawking temperature with event horizon relation reaches at critical point, $\beta_1=4.108$ with charge $q=0.1$. For $\beta<\beta_1$, only one extreme point exists, which means no phase transition will occur. For $\beta_1<\beta<\beta_2$, there are three extreme points on $T-r_+$ curve, we shall discuss whether this kind of black holes have first order phase transition or not below.

\begin{figure}[t]
    \begin{center}
        \subfigure[\;$\beta=5$]{\includegraphics[height=4.5cm]{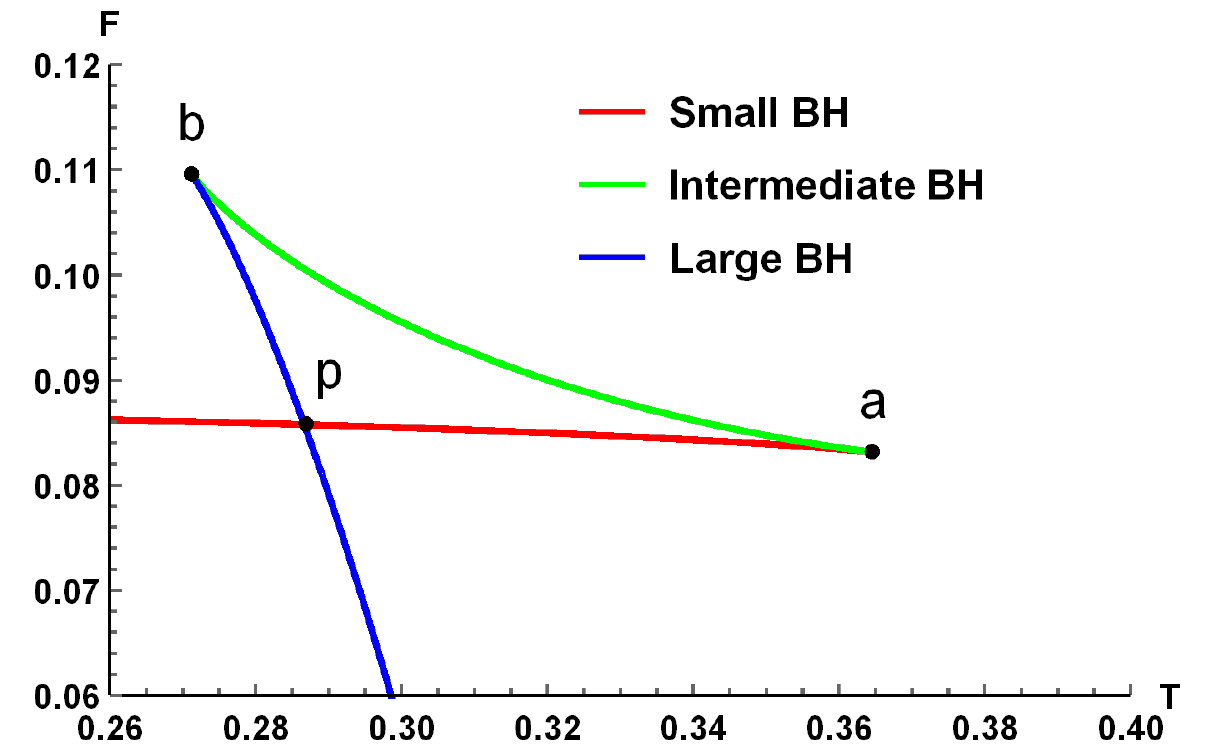}  
        \label{fig:fig2.a}}
        \subfigure[\;$\beta=4.7$]{\includegraphics[height=4.5cm]{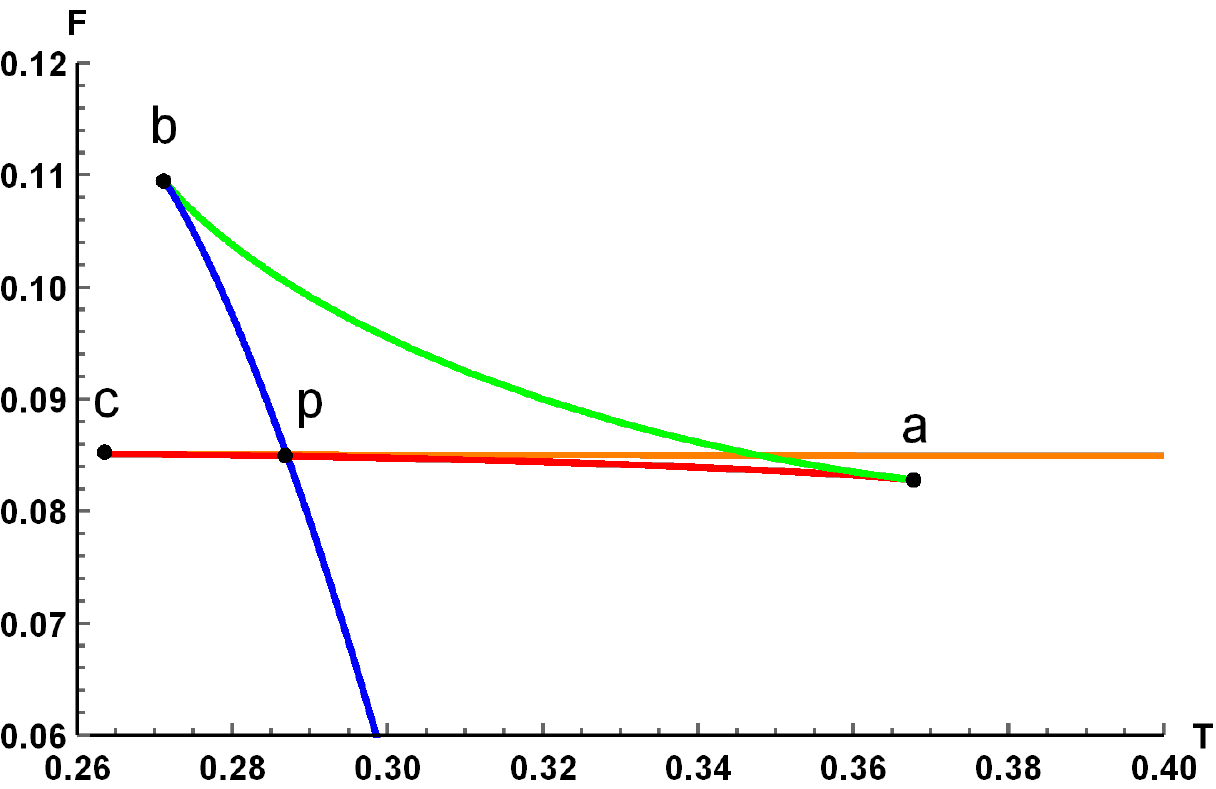}
        \label{fig:fig2.b}}\\
        \subfigure[\;$\beta=4.6$]{\includegraphics[height=4.5cm]{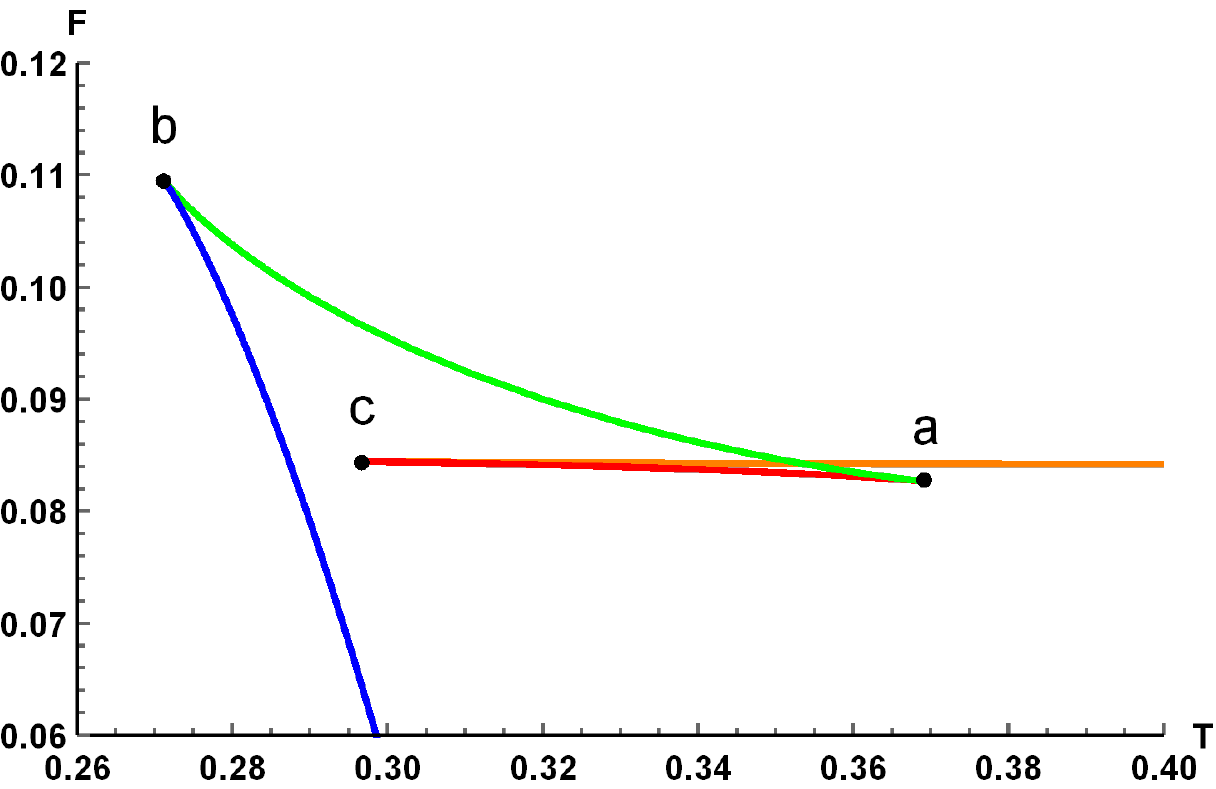}
        \label{fig:fig2.c}}
        \subfigure[\;$\beta=4$]{\includegraphics[height=4.5cm]{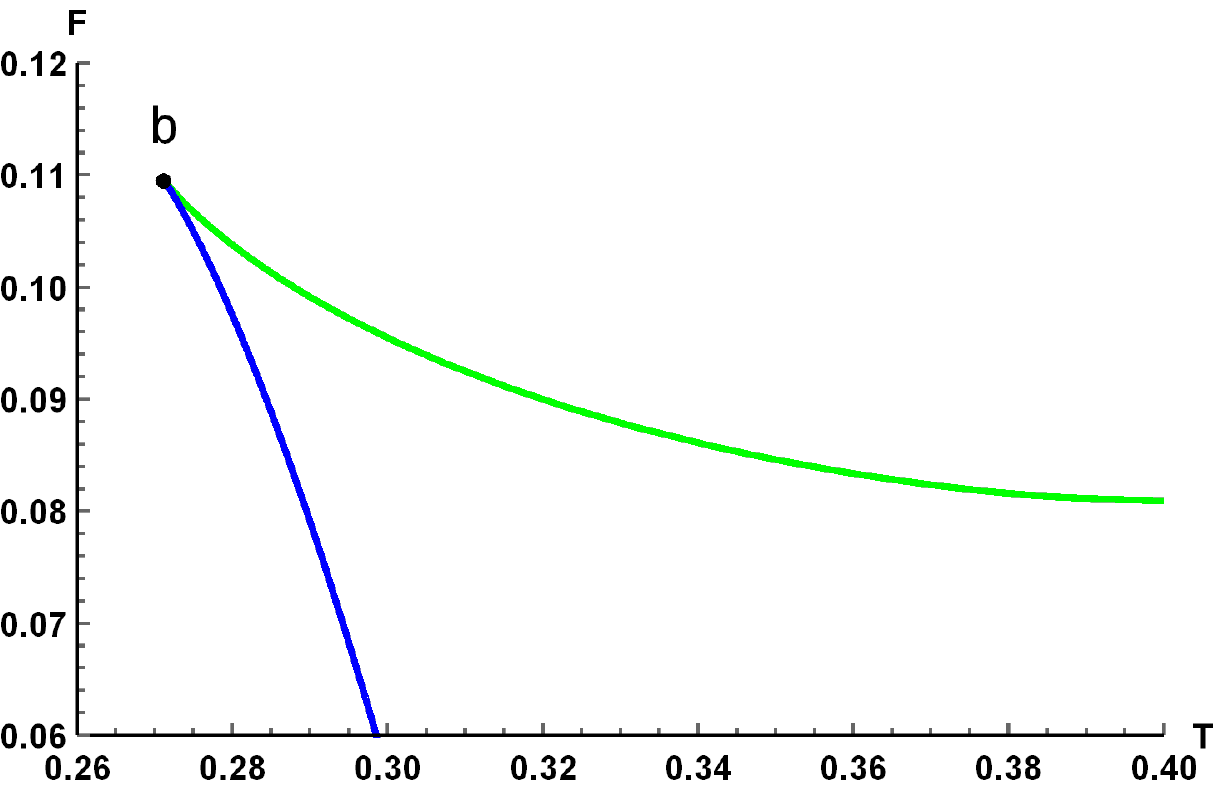}
        \label{fig:fig2.d}}
        \caption{\label{fig:Fig.2}Phase transitions $F-T$ diagram of Born-Infeld AdS black holes for different $\beta$, $\beta=5$ (a), $\beta=4.7$ (b), $\beta=4.6$ (c) and $\beta=4$ (d). (a) $\beta\geq\beta_2$ that the black hole phase structure is similar to RN AdS black holes. (b) $\beta>\beta_c$, the phase structure shows signs of disappearing. (c) $\beta<\beta_c$, the S-L phase transition disappeares. (d) $\beta<\beta_1$, the phase structures disappeares. Points $a$, $b$, $c$ shows the extreme points of $F$ and $T$, $p$ is the S-L phase transition point of the BI AdS black hole.}
    \end{center}
\end{figure}

To illustrate the black hole phase transitions, substitude Eqs. (\ref{eq:T}) and (\ref{eq:z}) into Eq. (\ref{eq:F}), and we can sketch free energy $F$ against Hawking temperature $T$ with different $\beta$ in FIG. \ref{fig:Fig.2}. It reveals that when $\beta>\beta_2$, there exists three phase structures, i.e. small BH, intermediate BH and large BH. The black hole in the section $p$-$a$-$b$-$p$ does not satisfy the stable equilibrium condition, thus is unstable. Besides, the small-large (S-L) phase transition is bound to occur accompanied by the jump of event horizon $r_+$ at the point $p$ in the figure, which is the first order phase transition point. S-L phase transition has been regarded as the reminiscent of liquid-gas phase transition for the Van der Waals fluid. When $\beta_1<\beta<\beta_2$, we have four black hole phases, for $T$ has three extreme points and two critical points. When $\beta_c<\beta<\beta_2$, since small BH and large BH has one intersection point, the phase transition still exists. When $\beta_1<\beta<\beta_c$, no S-L phase transition occurs. The $\beta_c$ is calculated as 4.636 for charge $q=0.1$. When $\beta<\beta_1$, there are only two solutions, and the black hole will tend to be the stable large BH state with no phase transition. When $\beta\rightarrow\infty$, the phase structure reduces to typical RN AdS black hole situation, which exactly corresponds to the former discussion.

\section{Lyapunov Exponents and Stability of Geodesic}
Lyapunov exponents are important for the study of chaotic systems. They can be used to measure the divergence and convergence rate of trajectories near the black hole. Lyapunov exponents in background metric have been explored \cite{Cardoso:2008bp}, which could describe the stability of geodesics around RN AdS black holes \cite{Guo:2022kio}. It is of our interest to describe timelike and null geodesics of particles around BI AdS black holes for the existence of nonlinear electrodynamical effect.

The effective potential combines the sum of multiple forms of energy except kinetic energy for particles around the black hole. To calculate the Lyapunov Exponents of particle orbits around black holes, the effective potential $V_r$ for radial motion is
\begin{gather}
\label{eq:Vr}
V_r=-\dot r^2,
\end{gather}
and begins with the equations of motion, the geodesic stability analysis in terms of Lyapunov exponents takes the form as \cite{Cardoso:2008bp}
\begin{gather}
\label{eq:lambda}
\lambda=\pm\sqrt{-\frac{V_r^{''}}{2\dot t^2}},
\end{gather}
is the Lyapunov Exponents for circular orbit on the equatorial plane, which represents the stability of circular geodesics in terms of the second derivative on $V_r$. Here the dot is the derivative with respect to proper time $\tau$ and the prime is the derivative respect to radius $r$. For simplicity, we can only take the $+$ symbol of right-hand side of Eq. (\ref{eq:lambda}) and refer to the Lyapunov exponent.

Massive particles and photons around BI AdS black holes travel under different geodesics. Massive particles travels the timelike geodesics in background metric around the BI AdS black hole. While photons propagate along null geodesics in the effective metric rather than the background metric due to the nonlinear electrodynamics effects \cite{Novello:1999pg,Kim:2022xum,Dittrich:1998fy,Shore:1995fz}. The two cases are discussed separately below.
\subsection{Timelike geodesics}
In order to obtain the expression of effective potential $V_r$ with respect to conserved quantities and orbit radius $r$, we should solve Hamilton canonical equation of the black hole. For circular orbit, the Lagrangian of geodesics on equatorial plane ($\theta=\pi/2$) is
\begin{gather}
\label{eq:L}
2\mathcal L=-f(r)\dot t^2+\frac{\dot r^2}{f(r)}+r^2\dot\varphi^2,
\end{gather}
here $\varphi$ is an angular coordinate. And generalized momentums are
\begin{gather}
\begin{split}
\label{eq:p}
p_t&=-f(r)\dot t=-E_n,\\
p_\varphi&=r^2\dot\varphi=L,\\
p_r&=\frac{\dot r}{f(r)},
\end{split}
\end{gather}
where $E_n$ is the energy and $L$ is the angular momentum of the particles. Those equations can be translated to 
\begin{gather}
\label{eq:varphi}
\dot\varphi=\frac{L}{r^2},\quad\dot t=\frac{E_n}{f(r)}.
\end{gather}
The Hamiltonian in terms of conserved quantities $L$ and $E_n$ satisfies
\begin{gather}
\begin{split}
\label{eq:H}
2\mathcal H=-\frac{E_n^2}{f(r)}+\frac{L^2}{r^2}+\frac{\dot r^2}{f(r)}=-1.\\
\end{split}
\end{gather}
Using the defination of effective potential $V_r$, we have
\begin{gather}
\label{eq:Vrr}
V_r=f(r)[1-\frac{E_n^2}{f(r)}+\frac{L^2}{r^2}].
\end{gather}

\begin{figure}[t]
    \begin{center}
        \subfigure[]{\includegraphics[width=0.5\textwidth]{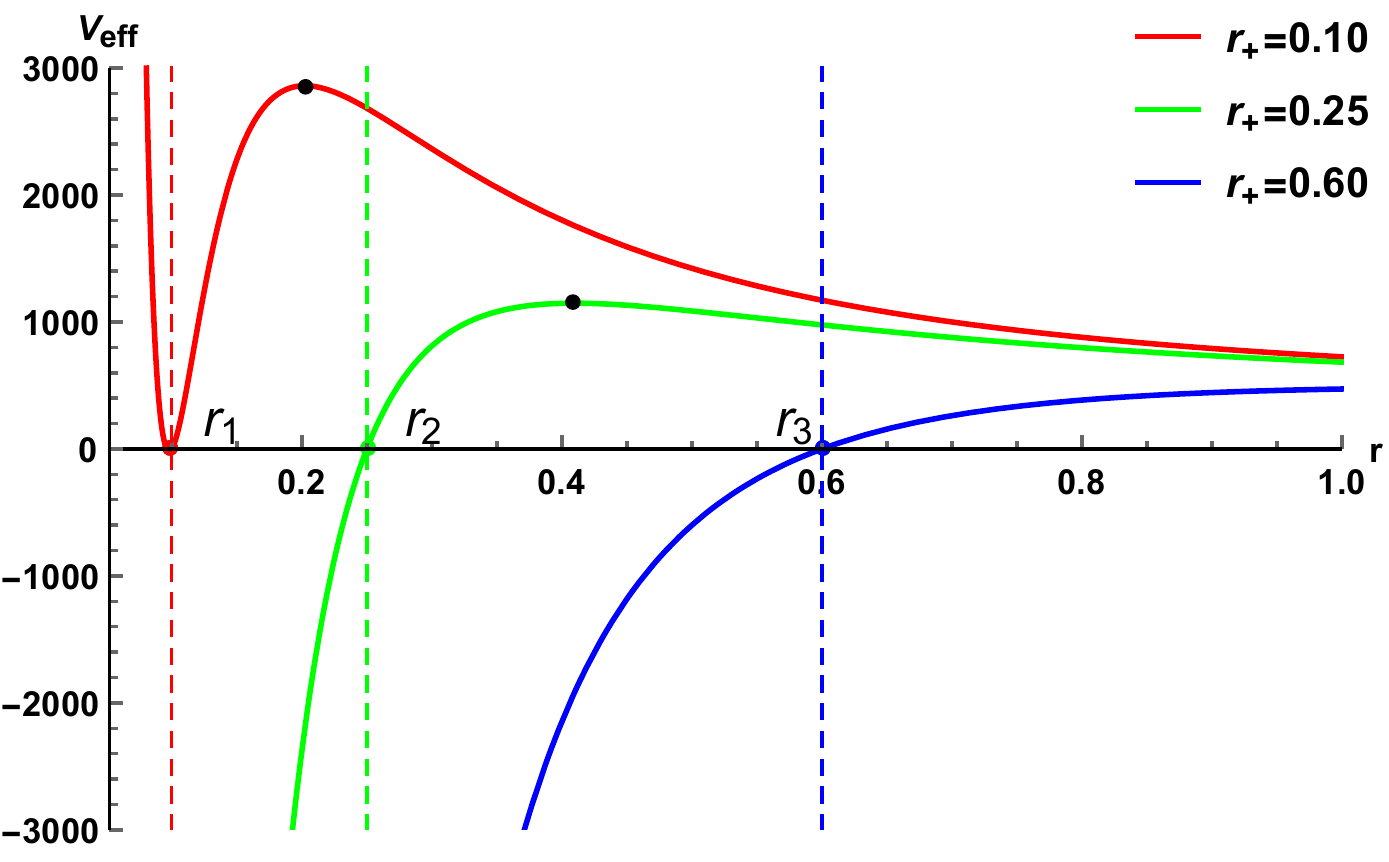}
        \label{ef}}
        \caption{\label{fig:Fig.3}Effective potential $V_r(r)$ of timelike geodesics as a function of $r$ with $q=0.1,\beta=50$. The dashed lines represent the radius of event horizon. $V'_r=0$ has no real solution of the blue curve for the curve has no extreme point (marked with black dots) unlike the other curves.}
    \end{center}
\end{figure}

To learn the features of timelike geodesics, we set $L=20l$ and $E_n=0$. Combining Eq. \ref{eq:MT}, we can see that $V_r$ is related to the BI AdS black hole parameter $q$ and $r_+$. The effective potential shows us the turning points and locations of stable or unstable equilibria, and $V_r-r$ relation is studied, as shown in FIG. \ref{fig:Fig.3}. We investigate the black holes with charge $q=0.1$ under different $r_+$ at 0.1, 0.25 and 0.6 corresponding to small BH, intermediate BH and large BH respectively. It is noteworthy that as $r_+$ increases, the maximum value of $V_r$ disappears, implying the disappearance of the unstability of the geodesics ($\lambda\rightarrow0$).

Now we can discuss the timelike geodesics Lyapunov exponent by the effective potential. Among the orbits of particles moving near a black hole, the most important one is the circular orbit on the equatorial plane corresponding to the black dot on FIG. \ref{fig:Fig.3}. For this orbit, it will be at a permanent turning point, and its radius has a certain value ($r=r_c$)  that satisfies $V_r(r_c)=V_r'(r_c)=0$ and $V''_r(r_c)<0$ for unstable geodecis \cite{Bardeen:1972fi}. 
Then, the energy and angular momentum can be expressed as
\begin{gather}
\label{eq:En}
E_n^2=\frac{2f(r_c)^2}{2f(r_c)-rf'(r_c)},\\
\label{eq:L2}
L^2=\frac{r^3f'(r_c)}{2f(r_c)-rf'(r_c)},
\end{gather}
which requires
\begin{gather}
\label{eq:ue}
2f(r_c)-rf'(r_c)>0.
\end{gather}

Combining the condition of timelike geodesics in Eqs. (\ref{eq:varphi}) and (\ref{eq:En}) to Eq. (\ref{eq:lambda}), one can get Lyapunov exponent
\begin{gather}
\label{eq:lt}
\lambda=\frac{1}{2}\sqrt{-(2f(r_c)-r_cf'(r_c))V''_r(r_c)}.
\end{gather}
According to Eq. (\ref{eq:ue}), we can see that $\lambda$ of timelike geodesics is real when the orbit is unstable which corresponding to $V''_r(r_c)<0$.

\subsection{Null geodesics}
In the BI NLED model, the dispersion relations of photons require revision due to NLED effects. Alternatively, the dispersion relations can be interpreted as those of the null vector with the geometry described by the effective metric \cite{Novello:1999pg,Kim:2022xum,Shore:1995fz}. This implies that the movement of light can be observed as electromagnetic waves traveling through a classical dispersive medium in a nontrivial vacuum \cite{Dittrich:1998fy}. The effective metric for static and spherically symmetric BI AdS black holes takes the form as \cite{He:2022opa}
\begin{gather}
\begin{split}
\label{eq:defc}
ds^2_{eff}&=(1+\frac{q^2}{\beta^2r^4})^\frac{1}{2}[-f(r)dt^2+\frac{1}{f(r)}dr^2+h(r)(d\theta^2+\sin^2\theta d\phi^2)],
\end{split}
\end{gather}
where
\begin{gather}
\label{eq:h}
h(r)=r^2(1+\frac{q^2}{\beta^2r^4}).
\end{gather}

Similar to timelike geodesics, the effective geodesic equations are given by a group of first-order Hamilton equations. Setting $\gamma$ to be the affine parameter, one can derive the effective geodesic equations as 
\begin{gather}
\label{eq:pm}
p^\mu=\frac{dx}{d\gamma},\quad \frac{dp^\mu}{d\gamma}+\Gamma^\mu_{\rho\sigma}p^\rho p^\sigma=0,
\end{gather}
where $\Gamma^\mu_{\rho\sigma}$ is the affine connection of the effective metric and $q_\mu\equiv G_{\mu\nu} p^\nu$ is dual vector of photons four momentum. For photons the conserved energy is $E_n=-q_t$ and angular momentum is $L=q_\phi$ like massive pariticles. Substitute the effective metric into the Hamiltonian for photons, we obtain
\begin{gather}
\label{eq:fe}
f(r)q^2_r-\frac{1}{f(r)}E_n^2+\frac{1}{h(r)}(q^2_\theta+L^2)=0.
\end{gather}
In the same way, we can also introduce the effective potential for null geodesics in effective metric
\begin{gather}
\label{eq:Ve}
V_r=f(r)[-\frac{E_n^2}{f(r)}+\frac{L^2}{h(r)}].
\end{gather}
Substitute this equation to the two prescriptions $V_r=0$ and $V'_r=0$, one can obtain
\begin{gather}
\label{eq:Ene}
\frac{E_n^2}{L^2}=\frac{f(r_c)}{h(r_c)},\\
\label{eq:ffe}
\frac{f(r_c)}{f'(r_c)}=\frac{h(r_c)}{h'(r_c)}.
\end{gather}

For null geodesics in effective metric, invoking Eq. (\ref{eq:Ene}) in Eq. (\ref{eq:lambda}) we can get the Lyapunov exponent for photons
\begin{gather}
\label{eq:ln}
\lambda=(1+\frac{q^2}{\beta^2r^4})\sqrt{-\frac{h(r_c)f(r_c)V''_r(r_c)}{2L^2}}.
\end{gather}
It can be described by Eq. (\ref{eq:ln}) that Lyapunov exponent of null geodesics is always real when the orbit is unstable, for $h(r_c)f(r_c)>0$ is assumed by the formula (\ref{eq:Ene}).

\section{Phase Transtions and Lyapunov Exponent}
Of interest is the black hole phase transitions, especially the S-L phase transition which is the reminiscent of Van der Waals fluid-gas phase transition, and various methods have been used to study them. Thermodynamical systems have the invariant free energy at the first order phase transition, so we can study the phase transition of black holes according to Maxwell's equal area law. It has been investigated that Lyapunov Exponent can characterize the RN AdS black hole's S-L phase transition \cite{Guo:2022kio}.  Next we extend this method to apply for massive particles and photons around BI AdS black holes.
\subsection{Timelike geodesics}
\begin{figure}[t]
    \begin{center}
        \subfigure[]{\includegraphics[width=0.6\textwidth]{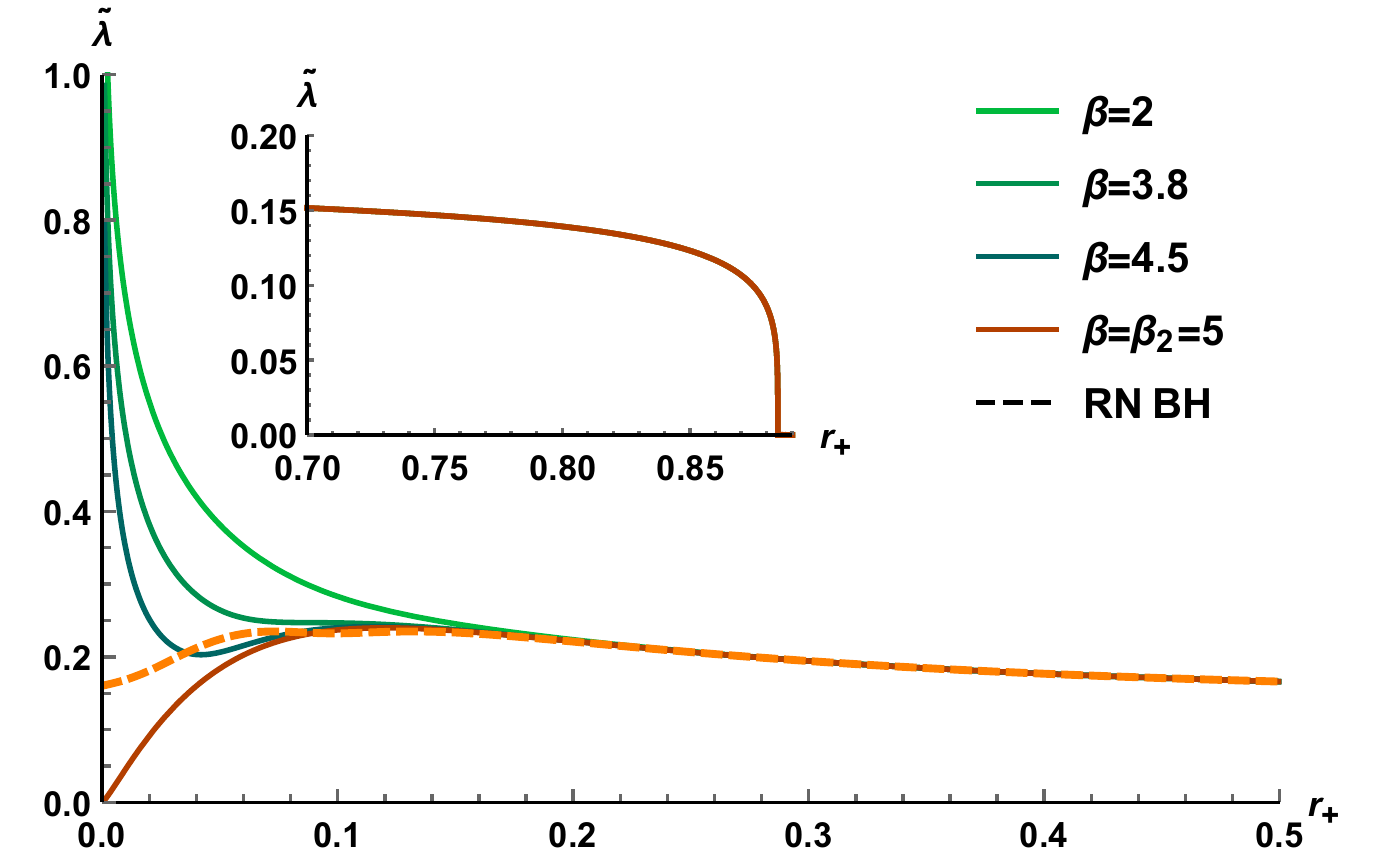}
        \label{lr}}
        \caption{\label{fig:Fig.4}$\tilde\lambda-r_+$ diagram for timelike geodesics in background metric with $q=0.1$, where $\tilde\lambda$ is taken as $\log_{100}(\lambda+1)$ (including the following). The behaviour of $\tilde\lambda-r_+$ with the changing of $\beta$ is qualitatively reminiscent of $T-r_+$ when $\beta\leq5$. The value of $\beta_2$ in $\tilde\lambda-r_+$ is unchanged compared with $\beta_2$ in $T-r_+$. Different from $T$, parameter $\tilde\lambda$ has one less extreme point as $r_+$ increases and $\tilde\lambda$ declines to 0 at $r_+=0.884$. When $\beta\rightarrow\infty$ the $\tilde\lambda-r_+$ relation reduces to RN AdS black hole type.}
    \end{center}
\end{figure}

In order to learn the relation between Lyapunov exponent and BI AdS black hole phase transition, we first study the relationship between event horizon and Lyapunov exponent of timelike geodesics in background metric.
To make it easier, plug the condition of timelike geodesics of Eq. (\ref{eq:L2}) into Eq. (\ref{eq:lt}),we have
\begin{gather}
\label{eq:lta}
\lambda=\frac{1}{2}\sqrt{-\frac{r_c^3f'(r_c)V''_r(r_c)}{L^2}}.
\end{gather}
Since the angular momentum $L$ is conserved, we can set it as $L=20$. Now we can see that Lyapunov exponent is closely related to radius of circular unstable geodesics $r_c$ which is determined by $V_r'(r_c)=0$. So the value of Lyapunov exponent $\lambda$ depends only on event horizon $r_+$ and charge $q$. It is intricate to solve this equatioin to reveal the relation between $\lambda$ and $r_+$ for different $\beta$ analytically, thus we include it in the numerical calculation and visualize the $\lambda-r_+$ relation in FIG. \ref{fig:Fig.4}.

Similar to FIG. \ref{fig:Fig.1}, there are also two branches in $\tilde\lambda-r_+$ relation of BI AdS black holes with fixed $q$, the threshold of which is defined by $\beta=\beta_2$ with $\lambda(r_+=0)=0$. We can regard $\beta>\beta_2$ black holes as RN AdS type and others as BI AdS type. Interestingly, the value of $\beta_2$ here is the same value of $\beta_2$ in $T-r_+$ and both of them satisfy $\beta_2=1/2q$. Lyapunov exponent can thus be seen as the analogy of Hawking temperature. Moreover, it is found that when event horizon $r_+$ increases, the Lyapunov exponent varnishes after a certain value, which means the unstable timelike geodesics in background metric won't exist around BI AdS black holes with over-larged event horizon. This phenomenon has been indicated by FIG. \ref{fig:Fig.3}, the extreme point of effective potential disappears when event horizon $r_+$ reaches specific value which corresponding to $V_r''=0=\lambda$.  In addition, as the increase of $r_+$, the value of Lyapunov exponent $\lambda$ almost has the same value with different BI parameters.

\begin{figure}[t]
    \begin{center}
        \subfigure[\;$\beta=50$]{\includegraphics[width=6.7cm]{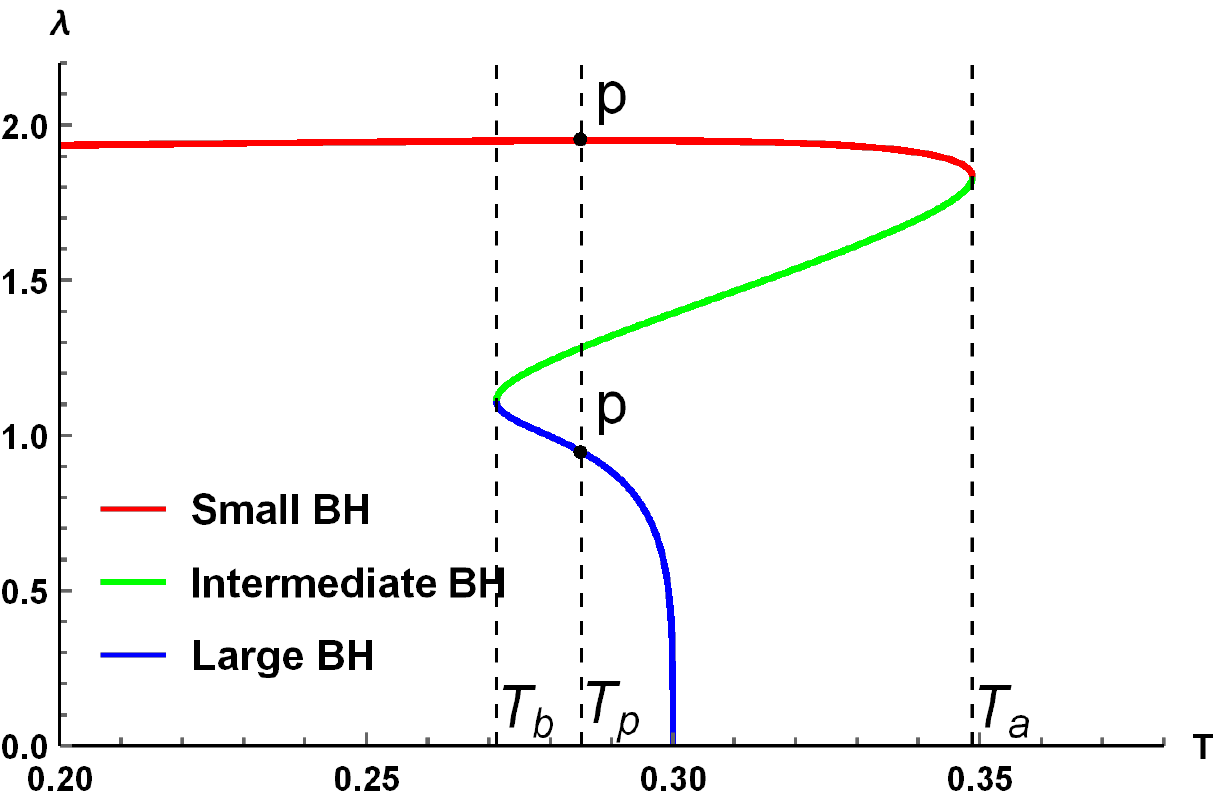}
        \label{fig:fig5.a}}
        \subfigure[\;$\beta=5$]{\includegraphics[width=6.7cm]{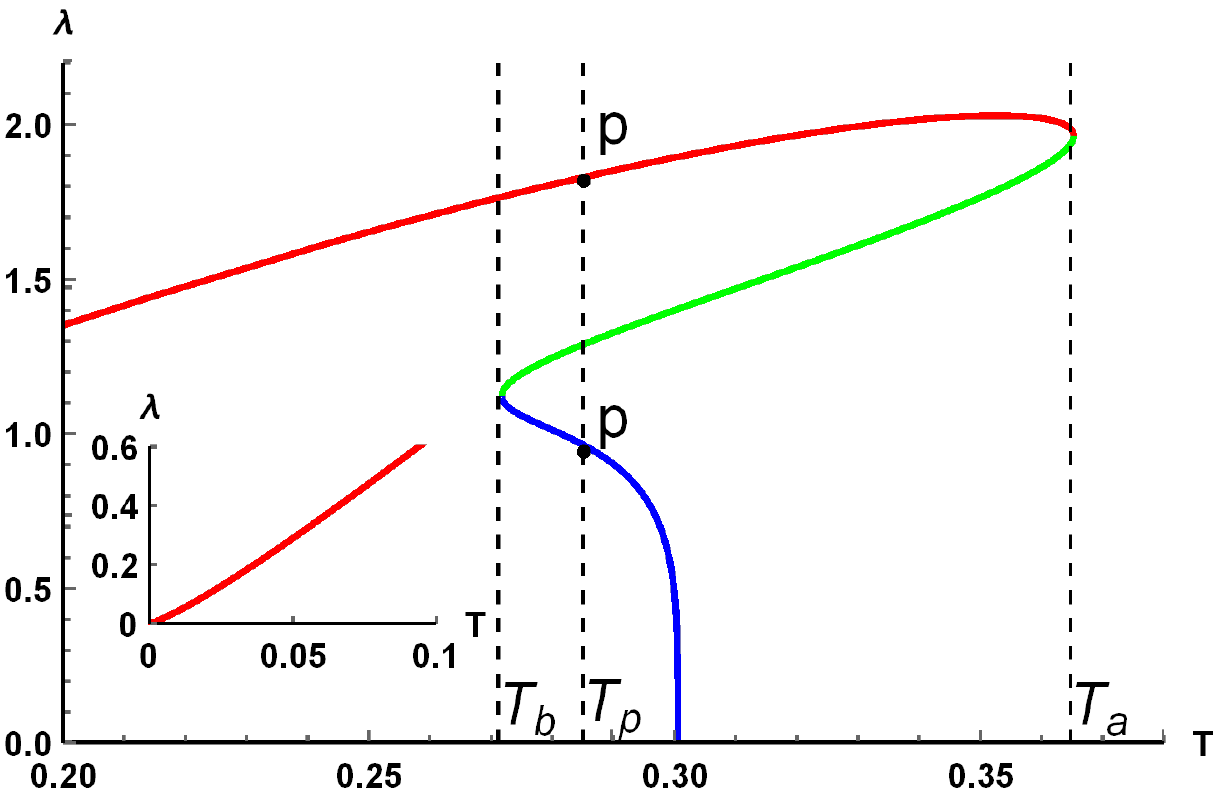}
        \label{fig:fig5.b}}\\
        \subfigure[\;$\beta=4.7$]{\includegraphics[width=6.7cm]{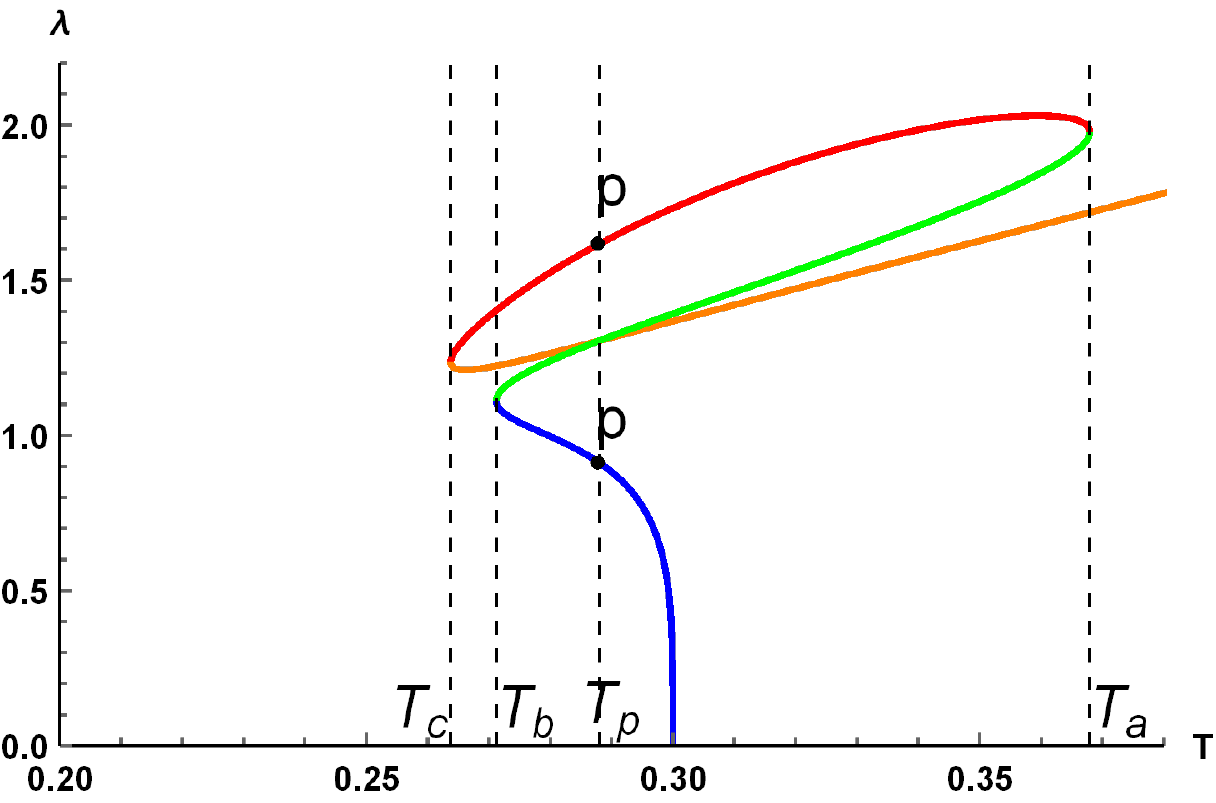}
        \label{fig:fig5.c}}
        \subfigure[\;$\beta=1$]{\includegraphics[width=6.7cm]{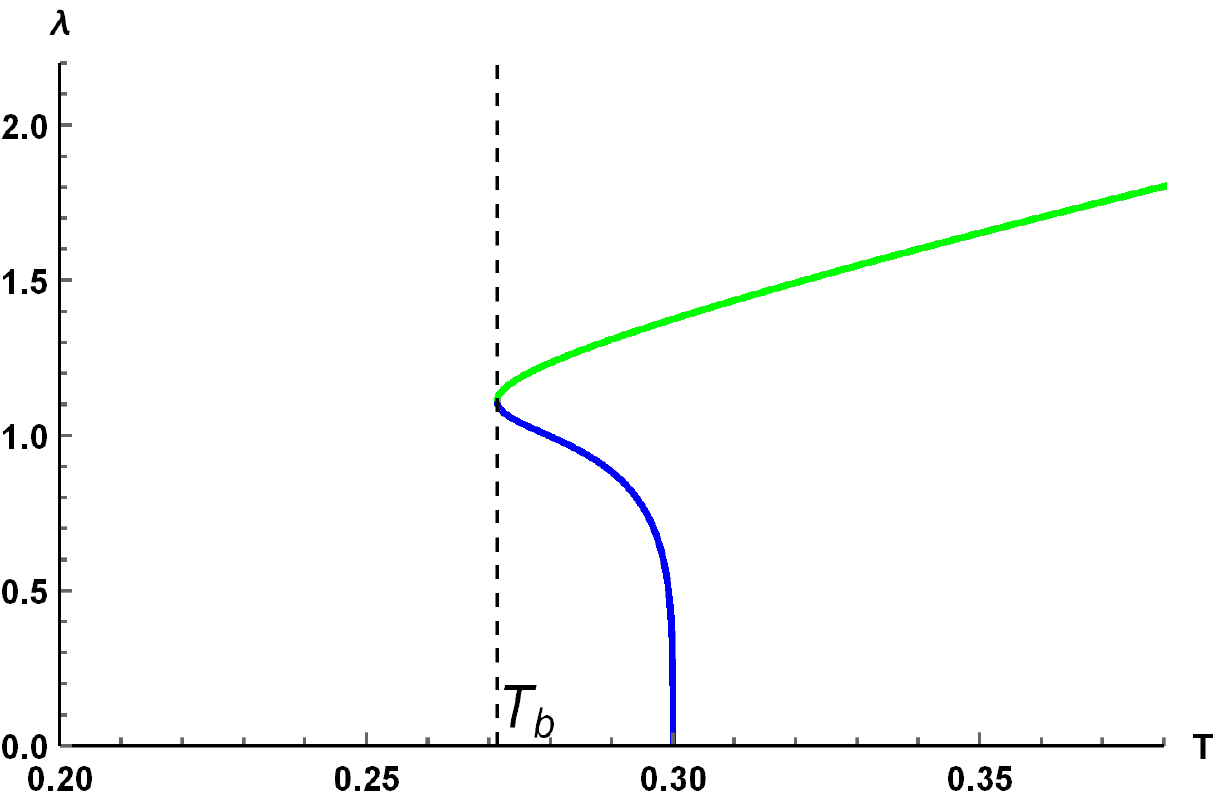}
        \label{fig:fig5.d}}
        \caption{\label{fig:Fig.5}Phase transitions $\lambda-T$ diagram of Born-Infeld AdS black holes for different $\beta$ with $q=0.1$. $\lambda$ is Lyapunov exponent of timelike geodesics in background metric. (a) $\beta=50$, the change of $\lambda$ in small BH is subtle. $\lambda$ will plunge to 0 as $r_+$ increasing. (b)  $\beta=5$, $\lambda$ will also decrease to 0 when $r_+\rightarrow0$ or $r_+$ is large enough. (c) $\beta=4.6$, except in large BH phase, it is similar to the massless geodesics. (d) $\beta=1$, $\lambda$ has no critical characteristics.}
    \end{center}
\end{figure}

We have derived Hawking temperature from the metric which is also the function of $r_+$ by Eq. (\ref{eq:T}). For the similarity between $T$ and $\lambda$, it's feasible to make Hawking temperature as a independent variable in the Lyapunov exponent phase transition diagram analogizing free energy. By substituting $T-r_+$ relation into Eq. (\ref{eq:lta}), we can get the relation of Lyapunov exponent $\lambda$ to $T$. Various S-L black hole phase transitions in different $\beta$ have been shown in FIG. \ref{fig:Fig.5}. To distinguish different phase structures of black holes (they are large, intermediate and small black holes respectively), we use different colors to mark them. Note that we depict the dashed line of Hawking temperature corresponding to the temperature of $a$, $b$, $c$ and $p$ points in FIG. \ref{fig:Fig.2}. The S-L phase transition points $p$ have also been marked.

The relationship between $\lambda$ and $T$ in the FIG. \ref{fig:Fig.5} is not just simply related by $r_+$. When BI parameter is large and close to RN AdS type, i.e. $\beta=50$, we find that for small BH phase, Lyapunov exponent is almost invariant, which indicates that Lyapunov exponent changes little with Hawking temperature $T$ (or $r_+$). Once the radius of the black hole event horizon increases to the S-L phase transition point $p$ corresponding to the Hawking temperature $T_p$, the Lyapunov exponent will decrease. When the Hawking temperature reaches at maximum $T_a$, the small BH turns to unstable intermediate BH in which $\lambda$ will change rapidly with $T$. And with the increase of event horizon $r_+$, $T$ will decrease to $T_b$ where intermediate BH transforms to the large BH. It is also predictable that Lyapunov exponent will plunge to zero as $r_+$ increases, which have been indicated in FIG. \ref{fig:Fig.4}. 

When $\beta=5$, $\lambda$ of small BH begins to decrease monotonously as $T$ decreases and the curve passes through $(0,0)$ exactly corresponding to $\beta_2$ curve in $\lambda-r_+$ and $T-r_+$ figure.  The phase transition can only occur at $\beta>\beta_0$, for it only has two phase structures of black holes and no phase transition for small $\beta$, i.e. $\beta=1$. Between $\beta_0<\beta<\beta_2$ we have another black hole phase marked by orange color to emerge. The Lyapunov exponent of small BH cannot reach zero and will tend to increase as $T$ decreases to $T_c$, i.e. $\beta=4.7$. For black holes with S-L phase transition, it is very useful to study critical behaviour of phase transition point by the property that $\lambda$ has a obvious difference between small and large phase. 

\begin{figure}[t]
    \begin{center}
       \subfigure[\;$\beta=50$]{\includegraphics[height=4.8cm]{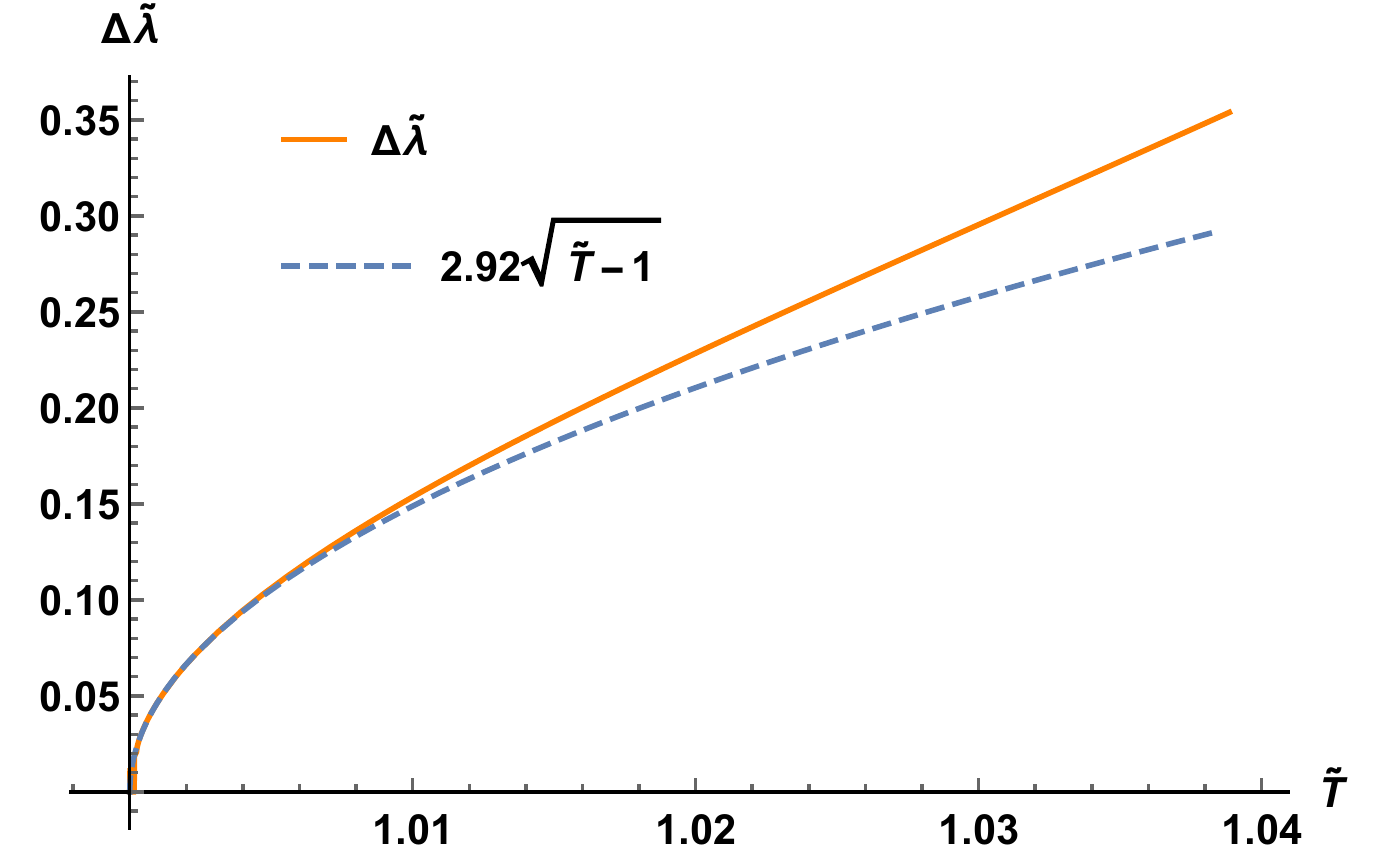}
       \label{fig:fig6a}}
       \subfigure[\;$\beta=5$]{\includegraphics[height=4.8cm]{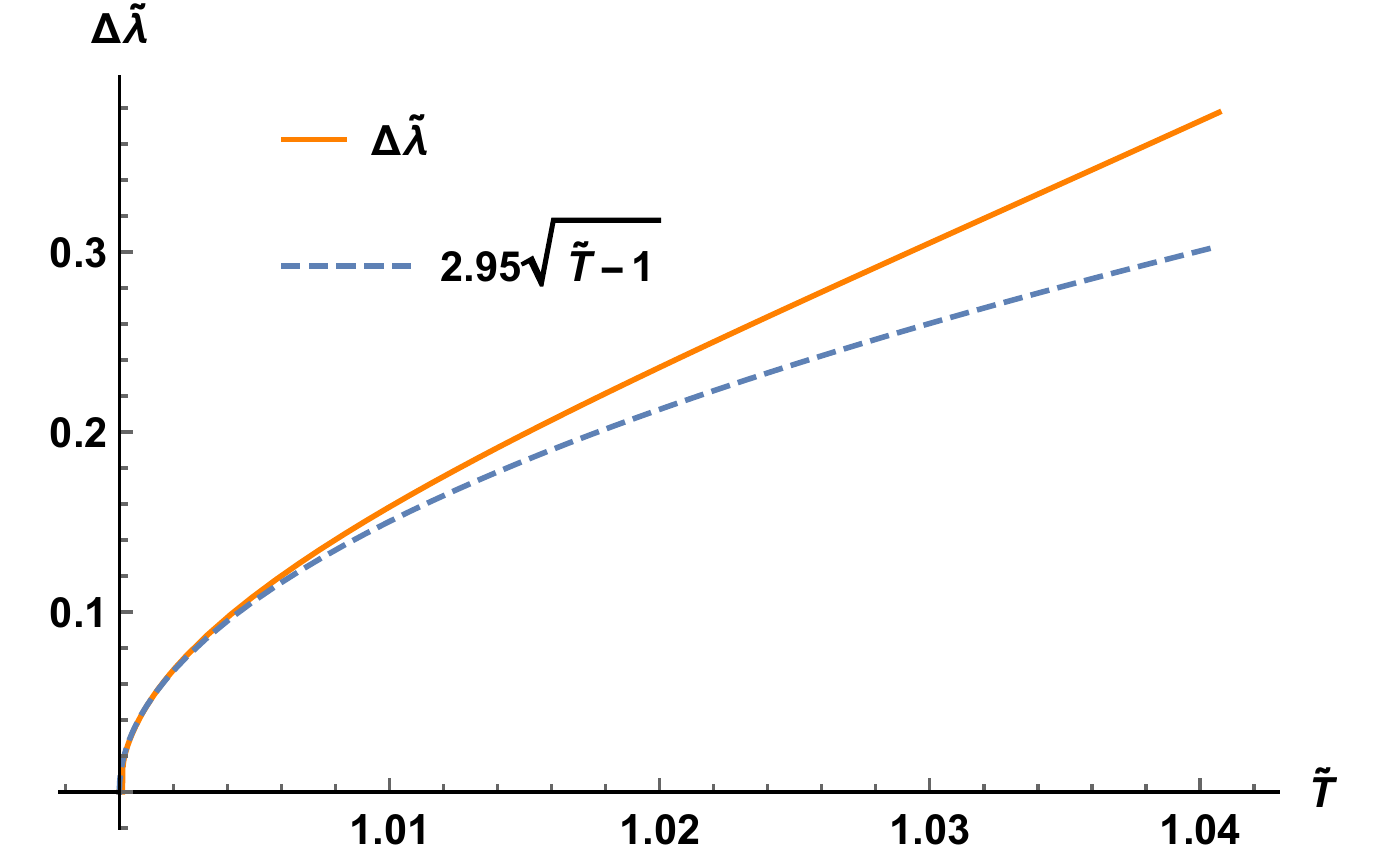}
       \label{fig:fig6b}}
       \caption{\label{fig:Fig.6} $\Delta\tilde{\lambda}$-$\tilde{T}$ figure with $C\sqrt{\tilde{T}-1}$ as reference function for timelike geodesics in effective metric. $\Delta\tilde{\lambda}=0$ in critical point as $\tilde{T}=1$. When $T$ is near the critical value $T_c$, $\Delta\tilde{\lambda}$ fit well with $2.92\sqrt{\tilde{T}-1}$ for $\beta=50$ and with $2.95\sqrt{\tilde{T}-1}$ for $\beta=5$.}
    \end{center}
\end{figure}

The phase structures is related to the slope of Lyapunov exponent. Besides, S-L phase transition can be described by the difference of Lyapunov exponent in BI AdS black hole phase transiton point $p$. At this point, the Lyapunov exponent is to be $\lambda_s$ for small BH phase  and $\lambda_l$ for large BH. In order to further analyze this connection, we study the change of Lyapunov exponent $\Delta\lambda=\lambda_s-\lambda_l$ at the first order S-L phase transition with different charge $q$. The Hawking temperature $T_p$ of the S-L phase transition changes with $q$. And we focus on the phase transition Hawking temperature $T_p$ to be near the critical temperature $T_c$. As $q=q_c$, two extreme points of $T-r_+$ will coincide and the phase transition vanishes where $T_p=T_c$ and $\lambda_s=\lambda_l=\lambda_c$. We have calculated several $T_c$ and $q_c$ in Table. \ref{table1} with different BI parameter $\beta$, and insert these quantities in Eq. (\ref{eq:lta}) to calculate $\lambda_c$. To learn critical behaviour of Lyapunov exponent for timelike geodesics around BI AdS black holes, we derive the relation of $\Delta\lambda/\lambda_c$ to $T_p/T_c$, as shown in FIG. \ref{fig:Fig.6}. For simplicity, we use $\Delta\tilde{\lambda}$ and $\tilde{T}$ to substitude $\Delta\lambda/\lambda_c$ and $T_p/T_c$ respectively. 

We know that when $\beta<\beta_c$, there is no second order phase transition structure, thus there is no critical temperature or critical behaviour when the BI parameter of BI AdS black holes is small.
For $\beta>\beta_c$, we find as $T_p\rightarrow T_c$, $\Delta\tilde\lambda$ shows a certain critical relationship with $\tilde{T}$. The critical exponents determine the qualitative behavior of a given system in the vicinity of its critical point. The critical exponent $\delta$ related to $\Delta\lambda$ is defined as,
\begin{gather}
\Delta\lambda(r_+)-\Delta\lambda(r_c)\sim |T-T_c|^\delta.
\end{gather}
To calculate this critical exponent $\delta$, we can follow the method of \cite{Banerjee:2012zm}. We can Taylor expand $\Delta\lambda$ near the phase transition point as 
\begin{gather}
\label{eq:tep}
\Delta\lambda(r_+)-\Delta\lambda(r_c)=\alpha|T_p-T_c|^{1/2},
\end{gather}
where
\begin{gather}
\label{eq:alpha}
\alpha=\left(\frac{1}{2}\left[\left(\frac{\partial^2 T}{\partial r_+^2}\right)_{q=q_c}\right]_{r_+=r_c}\right)^{-1/2}\left[\left(\frac{\partial\Delta\lambda}{\partial r_+}\right)_{q=q_c}\right]_{r_+=r_c}.
\end{gather}
We know that at the critical point $\Delta\lambda(r_c)=0$, and we can simplify Eq. \ref{eq:tep} as 
\begin{gather}
\label{eq:teps}
\Delta\tilde\lambda=k|\tilde T-1|^{1/2},
\end{gather}
here $k=\sqrt{T_c}\alpha/\lambda_c$.

That means the critical exponent $\delta$ of $\Delta\tilde\lambda$ at the critical point is 1/2 (the exponent on $\tilde{T}$ is 1/2), which is the same as order parameter in the van der Waals fluid \cite{Wei:2017mwc}. We calculate the coefficient $k$ for the case $\beta=5$ and $\beta=50$ and plot the curve on the FIG. \ref{fig:Fig.6} (blue dashed line). Specifically, for $\beta=50$, the parameter $k$ is numerically calculated as $k=2.92$ and for $\beta=5$, $k=2.95$. We can see that the results fit well with the $\Delta\lambda-T_p$ relation near critical point. The importance of this expression is that we can use another parameter to investigate the first order phase transitions between small and large phases of Born-Infeld AdS black holes. And conversely, we can use the temperature under the black hole phase transition to estimate value change of the Lyapunov exponent.

\subsection{Null geodesics}
\begin{figure}[t]
    \begin{center}
        \subfigure[]{\includegraphics[width=0.6\textwidth]{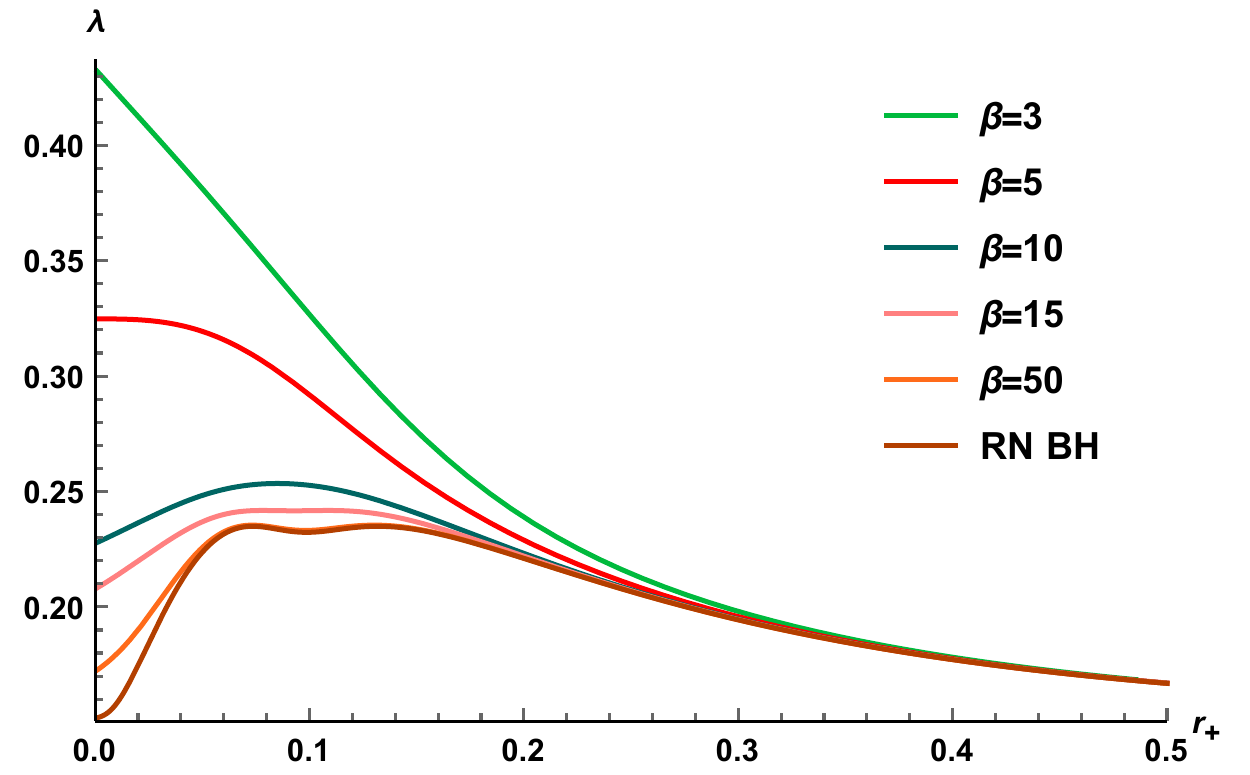}
        \label{lr2}}
        \caption{\label{fig:Fig.7}$\tilde\lambda-r_+$ diagram for null geodesics in effective metric. The value of $\tilde\lambda$ tends to go down as $\beta$ increases. When $\beta\rightarrow\infty$ the $\tilde\lambda-r_+$ curve reduces to RN AdS type. Parameters $\beta_1$ and $\beta_2$ have vanished in this case.}
    \end{center}
\end{figure}

For null geodesics in effective metric, we consider the nonlinear effect of BI AdS black holes, and study its thermodynamics by Lyapunov exponent. The Lyapunov exponent can be described by BI parameter $\beta$, charge $q$ and event horizon $r_+$. Analogy to the study of phase transitions in thermodynamics, we can investigate its relation with $r_+$. To learn how $\lambda-r_+$ develop with the change of $\beta$, we fix $q=0.1$, as shown in FIG. \ref{fig:Fig.7}.

We can see that for photons orbit along the null geodesics in effective metric, there is no threshold of $\beta=\beta_2$ to exist. Lyapunov exponent $\lambda$ monotonously decreases as event horizon $r_+$ increases for small value of $\beta$. And $\lambda$ for different $\beta$ converge to the same certain value with large $r_+$. Additionally, when $\beta\rightarrow\infty$ the $\lambda-r_+$ diagram still reduces to the situation for RN AdS black holes.

\begin{figure}[t]
    \begin{center}
    \subfigure[\;$\beta=50$]{\includegraphics[height=5cm]{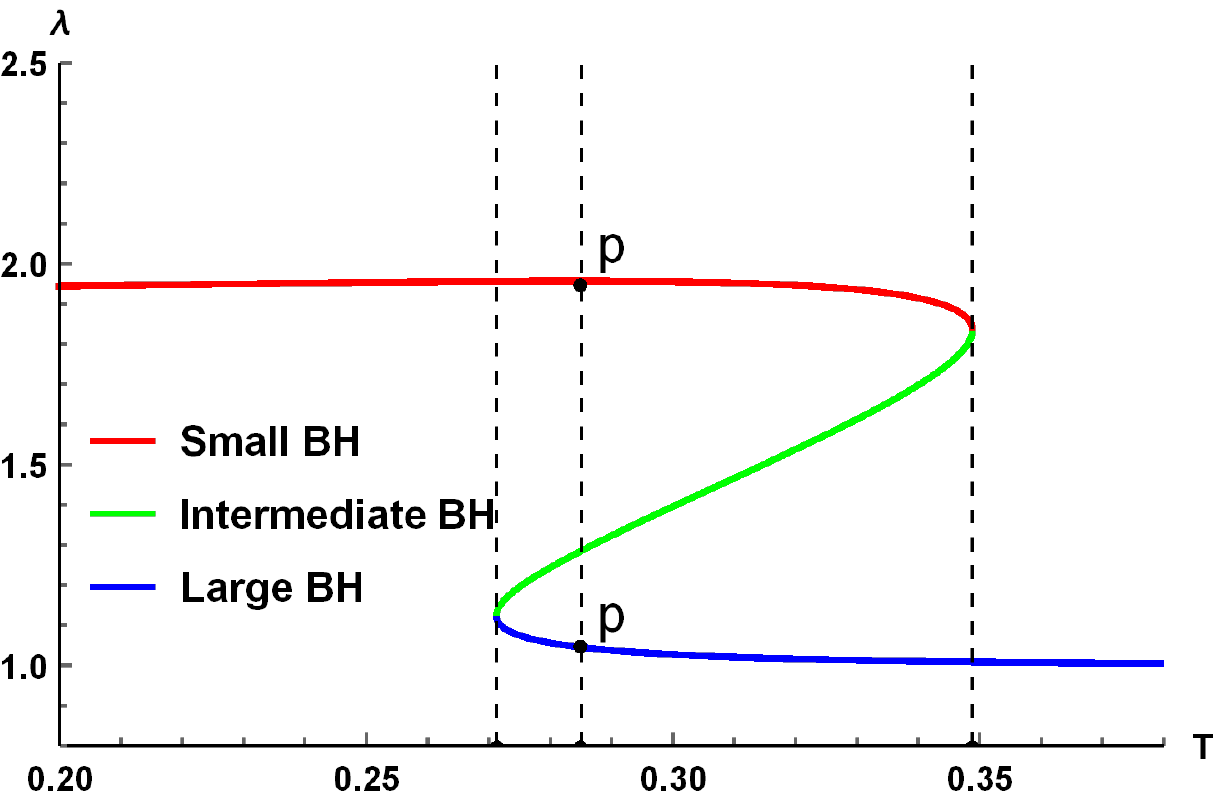}
    \label{fig:fig8.a}}
    \subfigure[\;$\beta=5$]{\includegraphics[height=5cm]{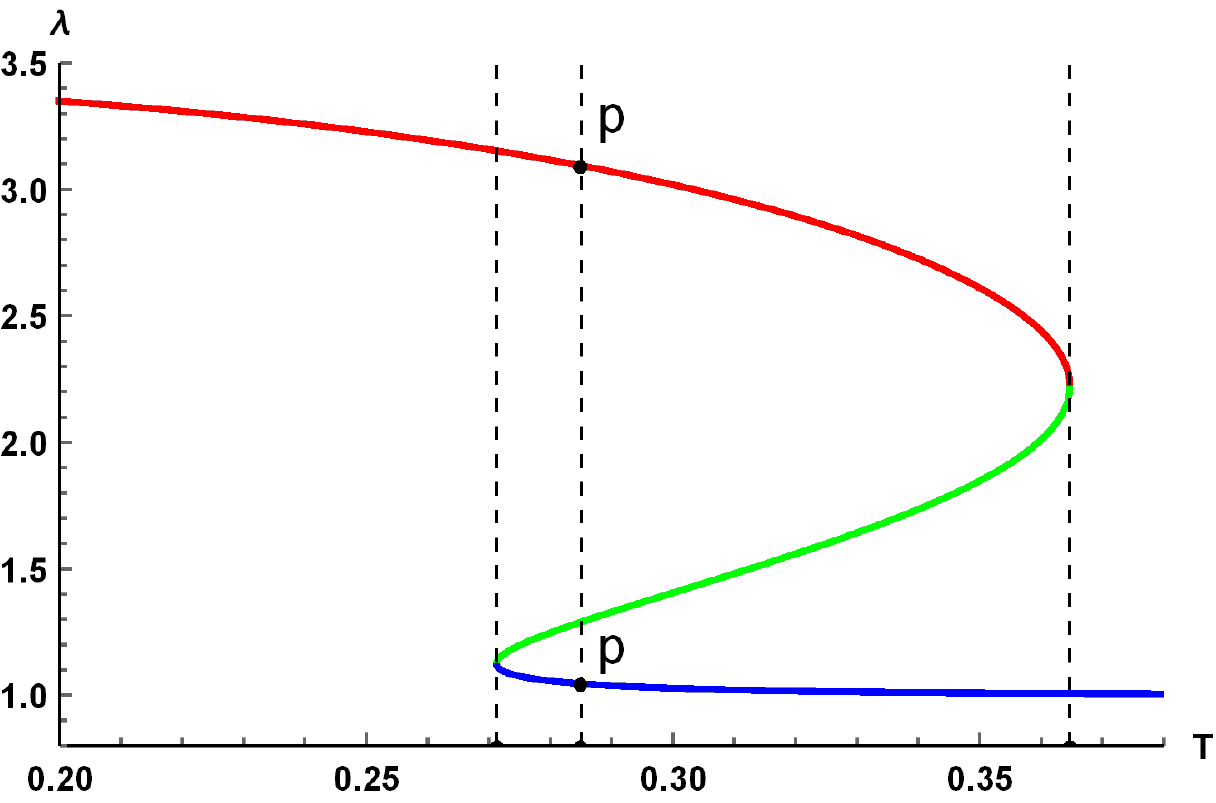}
    \label{fig:fig8.b}}\\
    \subfigure[\;$\beta=4.7$]{\includegraphics[height=5cm]{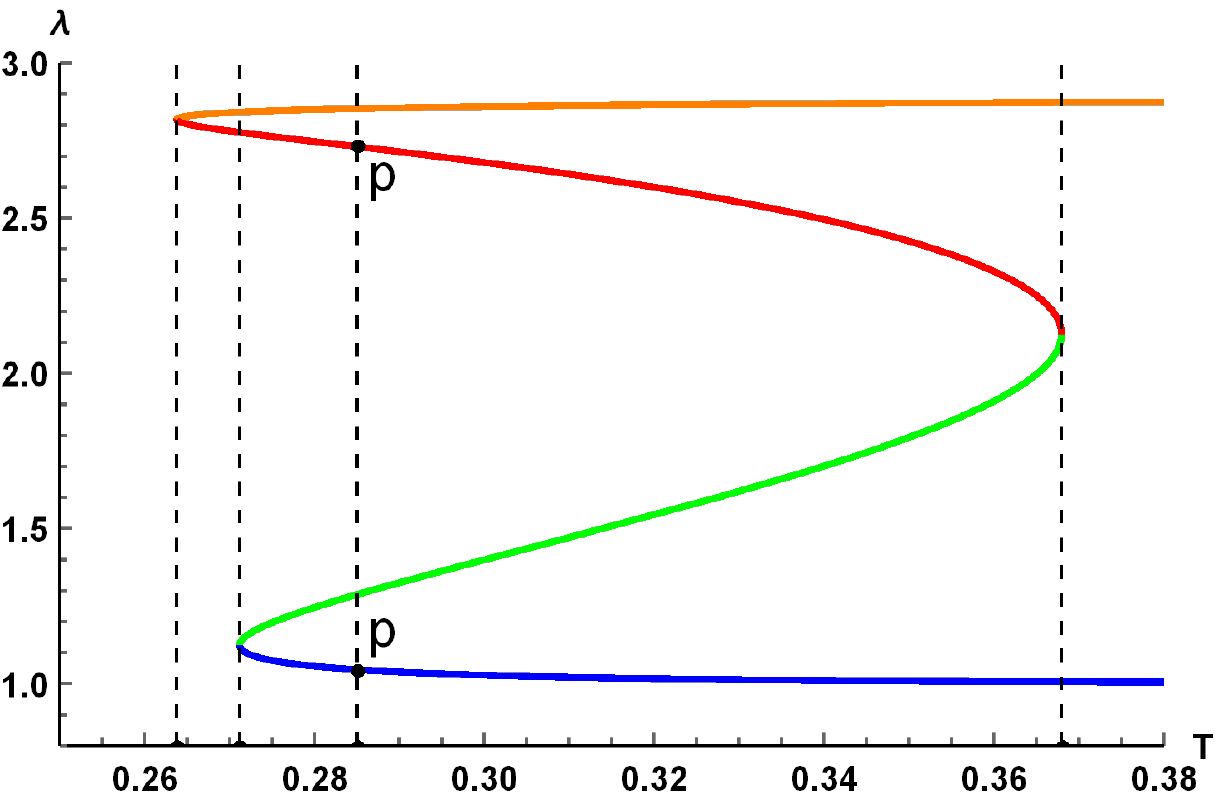}
    \label{fig:fig8.c}}
    \subfigure[\;$\beta=1$]{\includegraphics[height=5cm]{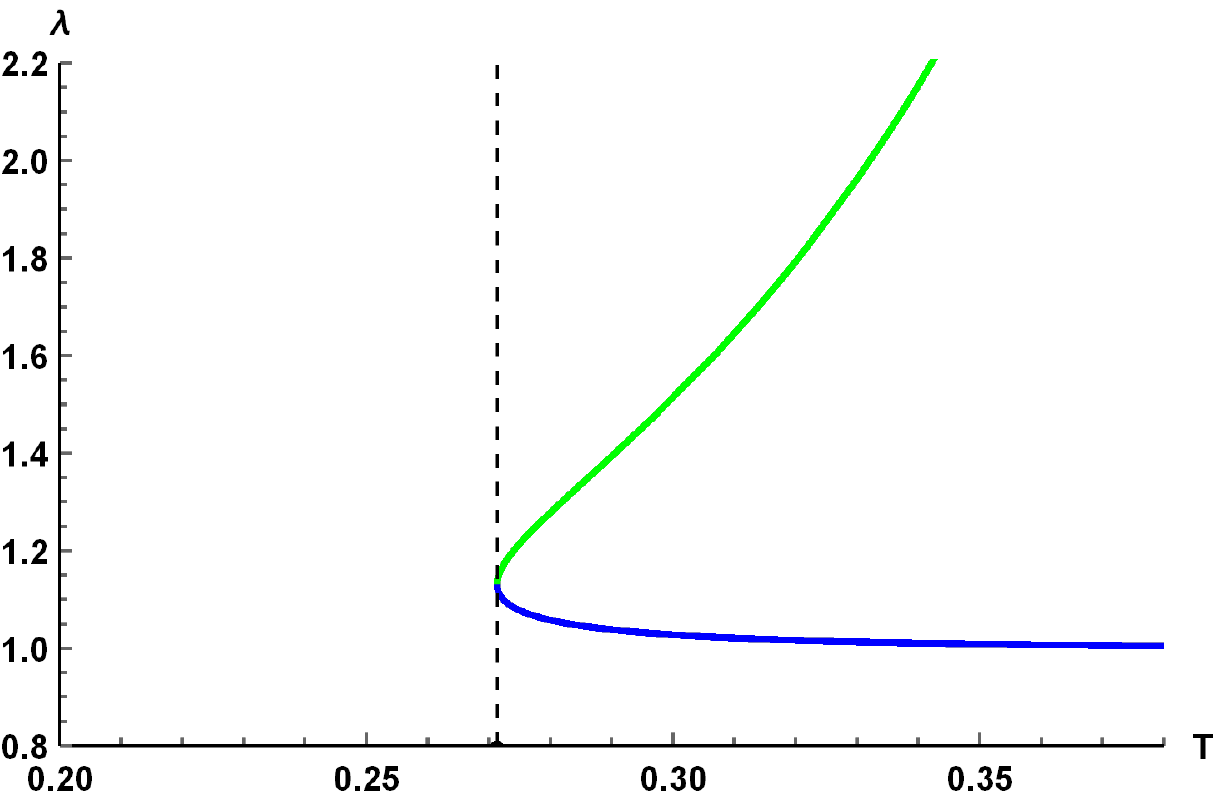}
    \label{fig:fig8.d}}
    \caption{\label{fig:Fig.8} Phase transitions $\lambda-T$ diagram of Born-Infeld AdS black holes for different $\beta$. Parameter $\lambda$ is Lyapunov exponent of null geodesics in effective metric. (a) $\beta=50$, $\lambda$ of small BH increases minutely as temperature increases. The phase transition from $p$ in small BH to $p$ in large BH  accompanies with the obvious change of $\lambda$.
(b) $\beta=5$, $\lambda$ monotonically decreases as the size of black hole increases.
(c) $\beta=4.6$, there is a new extreme point $T_c$. 
(d) $\beta=1$, $\lambda$ has no critical characteristics.}
    \end{center}
\end{figure}

It has been widely used of free energy to study BI AdS black hole phase structures, as shown in FIG. \ref{fig:Fig.2}. Here we use another parameter Lyapunov exponent $\lambda$ to simulate the BI AdS black hole phase transitions under null geodesics in effective metric in FIG. \ref{fig:Fig.8}. The strength of NLED effects can be characterized by parameter $\beta$.  In the limit $\beta\rightarrow\infty$ or $r\rightarrow\infty$, both the metric function in Eq. (\ref{eq:metric}) and the effective metric function in Eq. (\ref{eq:defc}) reduces to the RN AdS black hole. It is found that for large value of $\beta$ or $r_+$, the phase diagram respect to $\lambda$ and $T$ is qualitatively the same with RN AdS type \cite{Guo:2022kio}. Moreover, it is interesting to find that the nonlinear effect for null geodesics around the BI AdS black hole should be mainly expressed in small BH phase near $\beta=5$. It is found that for this small black hole (red line), the Lyapunov exponent tends to increase as the $T$ or $r_+$ decreases, where we have found the inverse trend for timelike situation in FIG. \ref{fig:Fig.5}. Especially for $\beta=4.7<5$ (BI AdS branch), another phase (orange line) emerges before small BH. The event horizon $r_+$ of this phase is smaller than small BH. And Lyapunov exponent $\lambda$ here tends to decreases as the $T$ or $r_+$ decreases, as shown in FIG. \ref{fig:fig8.c}. If $\beta$ is small enough, the small BH phase will disappear, as shown in FIG. \ref{fig:fig8.d}. For null geodesics on the BI AdS black hole with $\beta$ near $\beta=5$ and small $r_+$, it will exert significant nonlinear effect on Lypunov exponent. 

\begin{figure}[t]
    \begin{center}
    \subfigure[\;$\beta=50$]{\includegraphics[height=4.8cm]{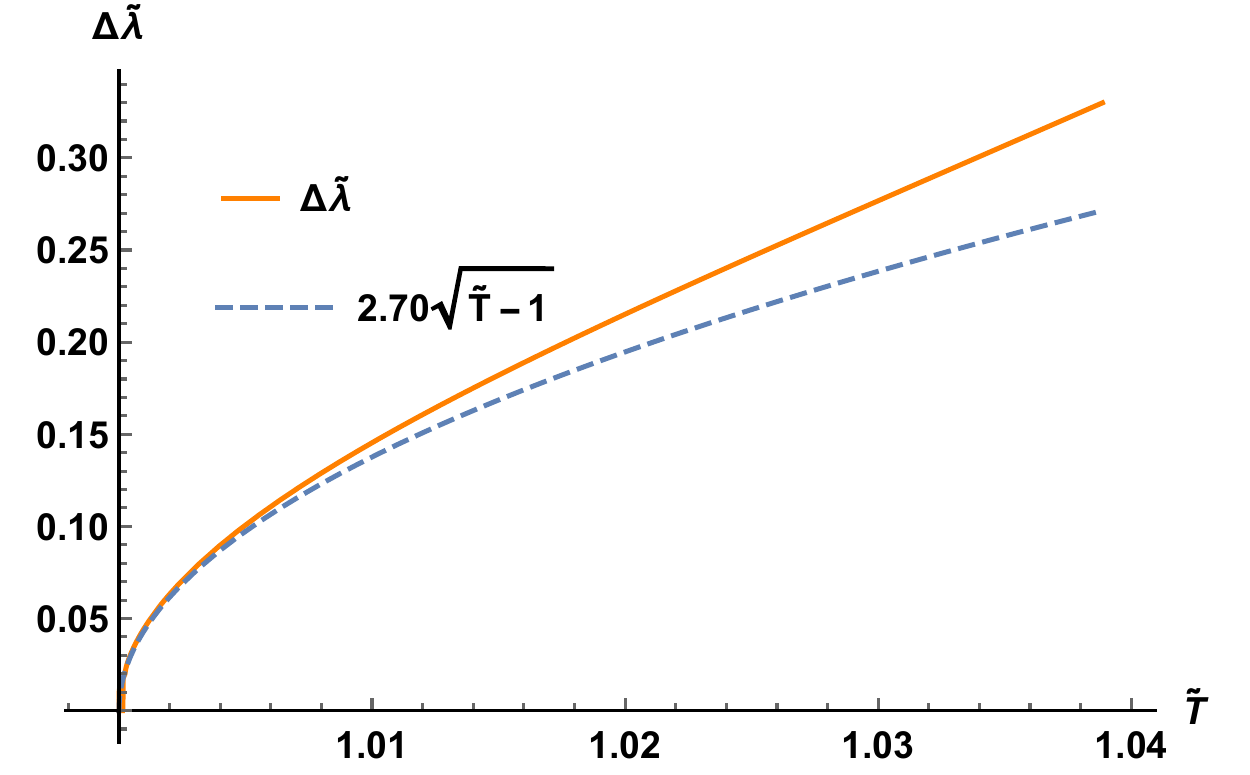}
    \label{fig:fig9a}}
    \subfigure[\;$\beta=5$]{\includegraphics[height=4.8cm]{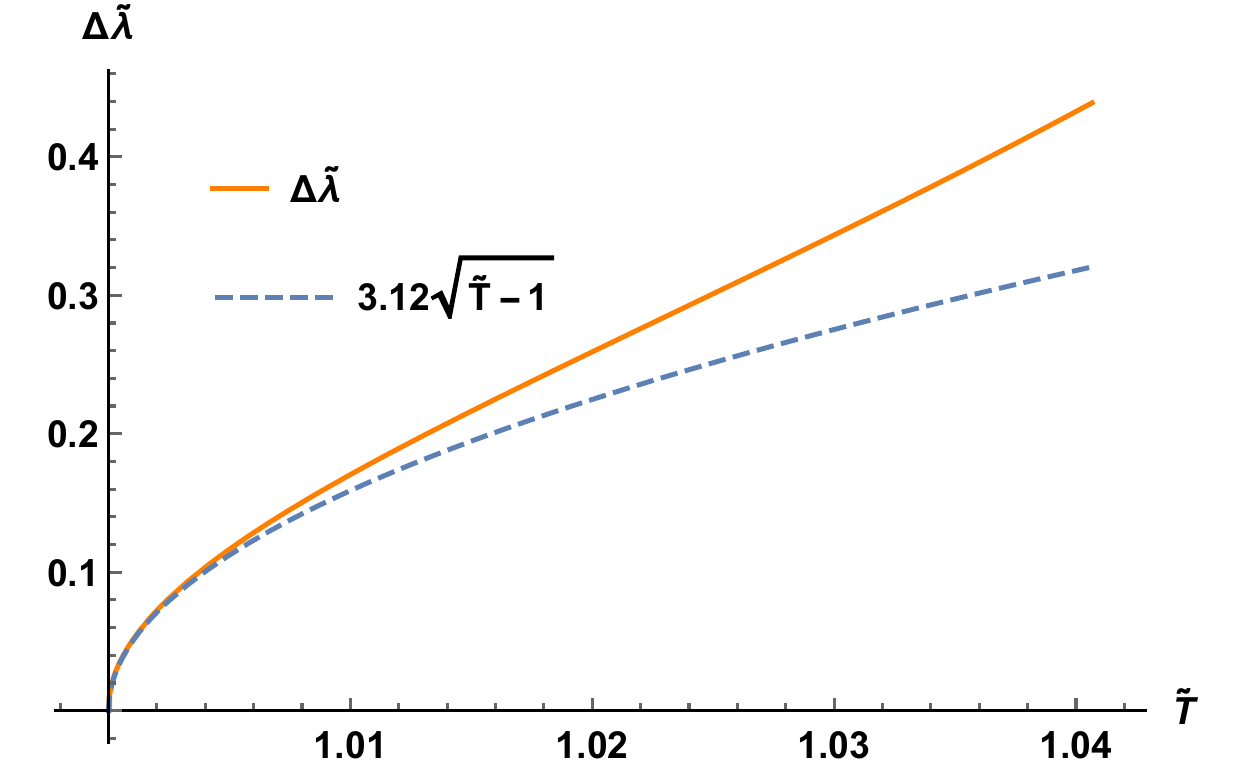}
    \label{fig:fig9b}}
    \caption{\label{fig:Fig.9} $\Delta\tilde{\lambda}$-$\tilde{T}$ figure with $C\sqrt{\tilde{T}-1}$ as reference function for null geodesics in effective metric. When $T$ is near the critical value $T_c$, $\Delta\tilde{\lambda}$ fit well with $2.70\sqrt{\tilde{T}-1}$ for $\beta=50$ and $3.12\sqrt{\tilde{T}-1}$ for $\beta=5$.}
    \end{center}
\end{figure}

Different from timelike geodesics, Lyapunov exponent $\lambda$ will tend to a certain value instead of plunging to zero as the increase of event horizon $r_+$ in large BH phase. And the first order S-L phase transition still exists in this kind of geodesics. Thus we can research phase transition of the null geodesics in effective metric once again by solving the timelike geodesics problem. To learn the critical behaviour near phase transition point for null geodesics, we depict $\Delta\tilde\lambda-\tilde T$ relation in FIG. \ref{fig:Fig.9}

Near the critical point, $\Delta\tilde{\lambda}$ and $\tilde{T}$ present critical behaviour for both $\beta=5$ and $\beta=50$. The critical exponent $\delta$ of $\Delta\tilde\lambda$ can be calculated like Eq. (\ref{eq:teps}) to be $1/2$. Thus, we can suppose that the critical exponent of $\Delta\tilde\lambda$ is independent of $\beta$ (if the black hole has a S-L phase transition) and geodesic type. We also numerically calculate $k$ for null geodesics, which are $k=2.70$ for $\beta=50$ and $k=3.12$ for $\beta=5$. $k\sqrt{\tilde{T}-1}$ still fit very well with $\Delta\tilde\lambda$ near $T_c$. It is also worth noting that our result is equivalent to the RN AdS type \cite{Guo:2022kio}. 

\section{Conclusion and discussion}
In this paper, we focus on exploring the relationships between BI AdS black hole phase transition and Lyapunov exponents $\lambda$. Then, the relations about Hawking temperature $T$ and event horizon radius $r_+$ for different BI parameter $\beta$ have been investigated. We have found that the different value of $\beta$ correspond to different number of black hole phase transition critical points. Factoring free energy derived from the first law of black hole thermodynamics, we have obtained three phase structures, i.e. small BH, intermediate BH and large BH respectively, and Hawking temperature $T_p$ of S-L phase transition point.

The main work of our paper is to characterize the black hole phase transition in terms of Lyapunov exponents for Born-Infeld AdS black holes. We discuss the context for timelike geodesics in background metric and null geodesics in effective metric respectively. For massive particles, by introducing the Hamiltonian we calculate the effective potential $V_r$ for timelike geodesics. Considering geodesic stability analysis in terms of Lyapunov exponents \cite{Cardoso:2008bp}, we have derived the expressions of Lyapunov exponent of particles around the circular orbit on black hole equatorial plane with respect to the $V_r$ which satisfy $V_r'(r_c)=0$. For photons in consideration of nonlinear effect around the BI AdS black hole, we use the effectice metric to derive the effective potential and the Lyapunov exponent for null geodesics around BI AdS black holes.

The relations of Lyapunov exponents and Hawking temperature for different $\beta$ with fixed $q$ analogying to the $T-r_+$ relations have been studied. When $\beta=\beta_2=(2q)^{-1}$, the Hawking temperature satisfies $\lim\limits_{r_+\to 0}T(r_+)=0$, which can be defined to be the threshold condition of these two branches. For timelike geodesics in background metric, the threshold value $\beta_2$ also exists in $\lambda-r_+$ diagram, which satisfies $\lambda(r_+=0)=0$. And Lyapunov exponents $\lambda$ will drop to zero as $r_+$ increases, which means BI AdS black holes will tend to be stable for large $r_+$. For null geodesics in effective metric, the relation of Lyapunov exponent $\lambda$ and event horizon $r_+$ has been studied and compared with timelike geodesics in FIG. \ref{fig:Fig.4} and FIG. \ref{fig:Fig.7}. Especially, we found the small BI AdS black hole near $\beta_2$ has significant nonlinear influence on photons. Besides, in order to figure out the intrinsic function of Lyapunov exponents change $\Delta\lambda$ in first order phase transition from small BH to large BH, we numerically plot $\Delta\lambda$ and fitting function in FIG. \ref{fig:Fig.6} and FIG. \ref{fig:Fig.9}. We also have expanded $\Delta\lambda$ and found out the critical exponent of $\lambda$ near the phase transition points to be 1/2. The theoretical analysis match well with the numerical results. 

Our research demonstrates the relationship between Lyapunov exponent and black hole thermodynamic phase transition in Born-Infeld AdS black hole. This research establishes a bridge between black hole thermodynamics (phase transition) and black hole chaos (Lyapunov exponent). And it probes the influence of nonlinear effect (effective metric) on Lyapunov exponent of null geodesics for BI AdS black holes which depends on the size of black holes and the BI parameter $\beta$. We hope that this work will bring new inspiration for black hole thermodynamics.

\begin{acknowledgments}
    The authors are grateful to Peng Wang, and Yadong Xue for useful discussions. This work is supported by NSFC Grant Nos. 12175212 ,12275183 and 12275184.
\end{acknowledgments}

\end{document}